\documentclass[manuscript,screen,acmsmall]{acmart}
\usepackage[toc,page]{appendix}
\usepackage{multirow}
\usepackage{tabularx} 
\usepackage[normalem]{ulem}
\usepackage{booktabs}
\AtBeginDocument{%
  }

\setcopyright{acmlicensed}
\copyrightyear{2027}
\acmYear{2027}
\acmDOI{XXXXXXX.XXXXXXX}

\acmConference[GROUP '27]{Proceedings of the 2027 ACM International Conference on Supporting Group Work}{2027}{TBD}

\author{Victoria Chui}
\author{Kelly McConvey}
\author{Shion Guha}

\affiliation{
  \institution{University of Toronto}
  \country{Canada}
}

\begin{document}

\title{A Sociotechnical Review of Algorithms in Health Systems: Technical, Cost, and Human-Centered Considerations}


\renewcommand{\shortauthors}{Chui et al.}

\begin{abstract}
Artificial intelligence (AI) applications in healthcare are becoming increasingly prevalent, to assist health systems, providers, and patients with tasks such as decision-making, risk prediction, and diagnosis. This increasing computational potential brings AI applications to the forefront of workplace decision making, often without full consideration of subsequent computational, organizational, and social costs. These applications are leveraged to reduce healthcare costs and increase efficiency of daily tasks, with model-related costs being considered at varying levels of granularity. To understand these trends, we critically analyze 114 papers to examine how cost-aware AI models have been developed for health systems. We explore the data, method, and outcome choices of these models, as well as their intersection with cost and human-centered concerns, highlighting the gaps in rigorous sociotechnical model design. From these trends, we define \textit{model costs} and subsequent dimensions, presenting insight into those studies reporting financial, computational, organizational and/or social measures. Further, we critique the benefits and challenges of evaluating model-related costs and sustainability concerns when developing AI models for health systems.
\end{abstract}




\maketitle

\section{Introduction}
Artificial intelligence (AI) models have been extensively used in health systems, and are increasingly implemented, for disease risk prediction, diagnosis, cost prediction, and decision-making assistance \cite{Erion2022,ACM_Gyldenkaerne}. 
As computational power and access to data-driven modeling continues to increase in health systems, AI (including predictive models) will further entwine health administrators, clinicians, and patients with both AI solutions' positive and negative impacts \cite{ACM_Gyldenkaerne, ACM_AlRamahi}. Due to expanding and aging populations, and increasingly expensive healthcare supplies, there has been greater interest in using AI models to predict care, resource, procedural, and insurance costs for patients upon arrival to health systems \cite{Elsevier_Salmons_1,Elsevier_Salmons_2,Elsevier_Orji_Ukwandu,RAND_Kapinos}. Prior GROUP literature has explored technological advancements impacting health care journeys, from communication networks in radiology \cite{pacs} to homecare devices \cite{patient_physician}. However, discussion on algorithmic sustainability and computational costs, alongside their implications for workplace decision making, is an increasingly prevalent yet unexplored topic.


In this paper, we consider a comprehensive scope of modeling-related costs by expanding beyond computational cost to consider processes such as data acquisition, model training, hardware and software requirements, power consumption, organizational burden, and required time and labour. We define this total cost as the \textit{model cost} of an AI application.
HCI scholars have similarly examined AI models involved in high-stakes environments such as homelessness \cite{Moon2024}, education \cite{McConvey2024}, and child welfare \cite{Saxena2022}. They have recurringly found that models for public-facing systems often do not consider the sociotechnical context of their end users, creating a disconnect between stakeholder needs and model solutions \cite{Moon2024,saxena_bureau,COMPASS}.
In this study, we examine how model design choices in current literature influence health systems, clinicians, patients, and AI developers by asking the following: 
\begin{itemize}
    \item \textbf{RQ1:} What combinations of methods, predictors, outcomes, and cost considerations are present in AI models for health systems, and how have these trends changed over time?
    \item \textbf{RQ2:} What model cost dimensions arise from these technical design combinations, and how can these concepts be operationalized in AI development?
    \item \textbf{RQ3:} How can researchers anticipate and mitigate concerns from algorithmic decision making in the workplace ?
\end{itemize}

By conducting a systematic literature review of cost-aware AI models in health system research from 112 peer reviewed papers and 2 white papers (published from 2005 to 2025), we examine the previous and current trends in model design. 
We present insights into which costs are being considered in these papers (AI as a reduction of financial cost, AI as a cost-effective alternative, or AI modeling related costs) and whether prioritizing a human-centered approach provides insight into each of these cost perspectives. Through interrogation of the technical design choices, cost considerations, and human-centered approaches of these models, we discuss the workplace impact of the distance between intended model implications and realistic sociotechnical settings. We expand past the notion that all AI models should have similar cost components to emphasize that claims of cost-effectiveness can shift labour, maintenance work, risk, and accountability across administrators, clinicians, patients, and technical staff. As cost reporting and framing is not standardized in AI literature, we provide guidelines and example measurements for cost dimensions relative to modeling in health systems. Our systematic literature search aims to contribute the following to the SIGCHI community: 

\begin{itemize}
    \item We discuss the patterns in technical model design and cost considerations of health system AI models. We elucidate those model characteristics that are most frequently associated with distinct model cost dimensions (\ref{sec6.1}).
    \item Derived from these trends, we develop a set of model cost dimensions, guidelines, and example measurements, to provide researchers with a preliminarily structured manner to surface a holistic cost assessment during AI development (\ref{sec6.2}).
    \item Mapping these dimensions onto trends detected in our corpus, we present human-centered design guidelines that consider cost, sociotechnical, and workplace contexts. We recommend developers to holistically consider and evaluate model costs through co-design with stakeholders, analyze the impacts of models in the workplace and on downstream stakeholders, and question the suitability of modeling (\ref{sec6.3}, \ref{sec6.NEW}, \ref{sec6.4}).
\end{itemize} 
We begin by presenting the background setting of AI models in health, relevant cost considerations, and human-centered algorithm design (HCAD) principles \cite{Baumer17}. We then discuss our methods for selecting and analyzing the literature, and explain our findings and subsequent wider implications.

\section{Background}

To place our work in AI and health, we examine previous literature from within and outside the SIGCHI community on AI models in healthcare and the current state of relevant cost considerations. 
Understanding the current landscape of biases considered by developers for AI models in health systems can illuminate design implications. Considering the costs of developing, implementing, and maintaining these AI solutions can further reduce tensions between expected model outcomes and realistic contributions to health systems. Emphasizing an HCAD approach \cite{Baumer17} can help designers create models that are more suited to each relevant health system, emphasize justice and fairness, more fully encapsulate sociotechnical context, and estimate modeling-related costs.

Other literature reviews conducted at the intersection of AI and health have looked intricately at explainable strategies \cite{ali_2023}, start-ups \cite{Zahlan_2023}, and sustainable goals \cite{Vishwakarma} for algorithmic applications \cite{ali_2023,Vishwakarma,Zahlan_2023}. 
Our review uses a similar methodology in its systemic nature and collection of approximately 100 papers; however, it differs with its focus on human-centered and cost considerations made by developers and stakeholders.

\subsection{Model Design Leading to Justice, Fairness, and Bias Concerns}

There has been a recent push within the SIGCHI community to consider bias and its implications throughout the modeling pipeline \cite{barnard_fairness,hort_bias}. Developers are increasingly shying away from black box methods, prioritizing explainability for collaborators, stakeholders, and end users \cite{IEEE_Massa,Elsevier_Mohanty,quttainah}. While models in healthcare can serve to efficiently automate decisions, particularly in resource allocation and diagnostic settings \cite{ACM_Kudyba,Xiao_automate}, certain demographics, populations, and stakeholders are often deprioritized and discriminated against through model results \cite{ACM_Pfohl,wojcik_discrimination}. This can include decreased model accuracy for certain subgroups (frequently defined by race, sex, ethnicity, disability, or age) \cite{rua_disability_RR,ACM_Pfohl}, an imbalance of data collection from subgroups and therefore underrepresentation \cite{giovanola_bias}, or privacy and security risks \cite{chen_security}. Researchers have previously explored how these types of biased models can result in underdiagnosis of specific subgroups (frequently Black patients) for skin cancer \cite{sangers_bias}, cost prediction \cite{obermeyer_bias}, or care allocation \cite{ledford_bias}. These unfair outcomes can be attributed to statistical, representational, societal, deployment, or other biases embedded in the AI pipeline \cite{giovanola_bias,hort_bias,ledford_bias,obermeyer_bias,sangers_bias}. Biases considered in our corpus include performance bias due to unrepresentative data \cite{ACM_Gyldenkaerne}, the implications of applying different fairness metrics and measurement error definitions \cite{ACM_Pfohl}, and (non)considerations due to age, race, insurance carrier, or socioeconomic status \cite{Elsevier_Davoudi}. For many of these studies, the data is representative of those patients actively seeking care, meaning only those with access and motivation to seek care will have their information recorded. This literature provides important context for the predictors, methods, and outcomes chosen by developers.

Implementing AI solutions in workplace settings where clinicians and patients analyze model results together (such as a general practitioner's office) can lead to power imbalances and disruption of the often fragile patient-physician relationship \cite{antoniak_NLP,Elsevier_Gentili,IEEE_Lee,patient_physician}. 
Gyldenkærne et al. [2020] push for a ``[participatory design] agenda'', allowing clinicians to contribute to the design and implementation of those ``technologies that affect their patient treatment and care, [introducing] a sensitivity to existing work practices and power relations'' \cite{ACM_Gyldenkaerne}. They found that establishing AI in health with the distinct aim of complementing patient and clinician needs expands the cognizance of electronic health records (EHRs) as data. Considering these fluctuating dynamics, combined with unravelling black box understandings of model workings, can encourage justice-centered AI solutions, reducing mistrust from stakeholders who are placed at a disadvantage due to AI implementation \cite{ACM_Pfohl,wojcik_discrimination}.

\subsection{Cost Considerations in Healthcare Modeling} \label{sec2.2}

AI models are often implemented in health systems to reduce spending on resources, diagnostic tests, patient care, and decision-making efforts \cite{kumar_costs}. 
Machine learning models capable of analyzing patient data - whether from electronic health records (EHRs), clinician notes, diagnostic images, prescriptions, or laboratory information - and providing treatment recommendations, diagnoses, or disease risk are increasingly used as cost-effective solutions to what usually requires intense administrative efforts \cite{RAND_Kapinos,JMIR_Rajkumar,waheed_costeff}. We expand the traditional definition of computational costs \cite{justus_cost,yu_cost} to define \textit{model costs}, which also includes the costs of acquiring data, validating, implementing, and powering models, environmental impact, and social and workplace costs (including time and labour) throughout the AI pipeline. Not all models will have the same composition of model costs, 
as the ``price'' associated with developing and maintaining models may include resources (medical supplies, testing, diagnosis), staffing needs, and digital infrastructure within health systems, and may or may not be financially quantifiable \cite{safehome}. We utilize the term ``model cost'' in our literature review to examine those models acknowledging what it takes to construct and use the AI. While cost prediction is one of the outcome categories derived from our coding, we separate this definition from model cost. We operationalize the term model costs into categories such as computational, financial (monetary), social, organizational, and time and labour costs (\ref{sec6.2}). Since these exact costs are not often reported in the literature, we do not provide exact measurements or quantitative comparisons between `high' and `low' cost models. We instead report patterns in technical model design choices that lead to consideration of different costs or human-centered principles.


The components of model costs that must be considered will differ depending on each project's needs, as costs and cost-awareness manifest uniquely in distinct health settings. It should then be considered to whom these costs are relative and who is responsible for them, whether an individual organization or multiple parties. For AI developers, each model cost should be clearly outlined and explicitly mentioned in development conversations. Reducing healthcare costs is a frequent motivation for AI implementation, as these solutions are often branded as ``cost-effective;'' allocating resources and analyzing patient care efforts and treatments more efficiently than human counterparts. However, they lack a consideration of burdens on the health systems, administrators, clinicians, and patients \cite{implications_elendu}. Cost-awareness in hospital settings is influenced by both internal workers' professional knowledge and prior experience as well as external patient financing policies \cite{financing}. Health clinics administer funds and resources in a separate manner to hospitals (not operating in-patient services), while still allocating the majority of funds to clinician and staff salaries. However, funding patterns differ depending on administrative organization (such as group or shared clinics) \cite{clinic_funding}. Due to the diversity of digital and e-Health applications, their cost-effectiveness can be a complex calculation, with researchers calling for further efforts into streamlined implementation and evaluation \cite{digital_eHEALTH,eHealth2}. 
Considering the costs of data acquisition, model training, execution, and implementation can ensure AI-related spending aligns with current bureaucratic processes and complements their infrastructural procedures \cite{saxena_bureau}.

Sub-components of model costs have previously been initiated, often through the definition of computational costs \cite{model_cost_application_1,model_cost_application_2,model_cost_application_3,model_cost_application_4}. Researchers have considered computational complexity and related costs to influence their choice of sampling techniques \cite{model_cost_application_1}, analyze cognitive effort and behaviour \cite{model_cost_application_2}, and choose a quantitative analysis method \cite{model_cost_application_3,model_cost_application_4}. Expanding on computational costs, models also need to consider the time and labour associated with implementation and upkeep (often an organizational burden), as well as the impact on human roles and relationships (physician-patient relationships, equity impacts - social burdens). Understanding how these costs become embedded in data derivation and AI pipeline stages can help developers more explicitly state how cost-effective these solutions are for all stakeholders \cite{Elsevier_Li,Elsevier_Mohanty}. Additionally, the absolute values of any reported monetary costs cannot be directly compared to 2026 values, without a proportional currency re-evaluation. While our corpus spans 2005-2025, we do not compare direct costs for this reason, instead choosing to examine the trends in how model costs are acknowledged and treated, independent of their absolute monetary value. Future work will supplement our analysis of nominal counting of cost considerations to synthesize the magnitude of model cost values, ensuring standardization across currencies and years.


\subsection{Human-Centered Algorithm Design} \label{sec2.3}

AI research across multiple disciplines has implemented Baumer's human-centered algorithm design (HCAD) framework \cite{Baumer17} to extrapolate trends in design, development, and implementation choices for public-facing AI \cite{Saxena2022,mcconvey2023,Moon2024,Chui2023}. Previous SIGCHI work has highlighted design decisions made by developers, policy makers, users, and commissioners in child welfare \cite{Saxena2022}, social media \cite{social_media}, education \cite{mcconvey2023}, and homelessness \cite{Moon2024}. 
Human-centered design strategies in healthcare AI have previously included focus group discussions with end users \cite{rwanda,thieme}, policy reviews to understand organizational priorities \cite{rwanda}, co-design sessions to elicit stakeholder feedback \cite{rwanda,thieme,ting}, and iterative stakeholder conversations to understand relevant sociotechnical context \cite{PubMed_Soliman,hcds_book,Chui2023}, as also applied to digitized healthcare \cite{smart_health} and e-Health \cite{e-health}. 
Deploying these strategies can assist with mitigating biased model outputs by understanding the process(es) behind data collection and working past (potentially dangerous) modeling assumptions \cite{hcds_book,Chui2023}.

Outside of health, Saxena et al. found invisible power dynamics for workers in child welfare systems, that would not be detected or considered without human-centered methodologies \cite{Saxena2022}. Similar tensions between model promises and realistic deliverables have manifested in power structure challenges \cite{home_health}, risk assessment biases for demographic subgroups \cite{ledford_bias,obermeyer_bias}, and privacy and security risks \cite{chen_security}. There is a need to address the biases prevalent in health system AI decisions and applications, as well as understand the sustainability of model cost and the efforts made to reduce healthcare spending with cost-effective alternatives.

Half of the papers in our literature corpus (57/114) compare the performance of multiple models (such as neural networks, decision trees, and random forests) for the same purpose (\ref{sec5.1}).
In an increasingly model-agnostic world, cost-aware AI is crucial in order to follow the evolution of human-centered AI. Encouraging developers to divest from selecting only the quantitatively ``best'' performing model, towards fair and equitable solutions developed with stakeholders, human-centered AI must increasingly prioritize cost-aware conversations \cite{cost_ethicalAI}. 
To temporally examine the proportion of papers leveraging human-centered principles, we evaluate the \textit{theoretical} considerations of contextual information, \textit{speculative} extrapolations of innovative model choices and, \textit{participatory} involvement of stakeholders in our corpus \cite{Baumer17} (Table \ref{table:2}, \ref{sec6.3}). 

\section{Methods}

\subsection{Scoping Criteria}

AI models in our context include those models utilizing a machine learning or inferential statistic technical basis for prediction tasks. We only include those models with clear methodologies, input data, and evaluation metrics specified either in the body of the paper or supplementary methods. To contribute to a health system, the models need to have tangible applications for a health department or administration, with intended end users being health system workers, clinicians, or patients. We center our review on clinical health applications in hospitals or related clinic departments, as opposed to modeling for preventative medicine, home care, or external mental health support. The review criteria are listed below. To identify AI models of interest in this context, we used search terms including: ``health care,'' ``healthcare,'' ``health,'' ``AI,'' ``artificial intelligence,'' ``ML,'' ``machine learning,'' ``cost,'' ``computational cost,'' ``model,'' and ``algorithm''. The combinations of search terms and full queries per database can be found in Appendix \ref{appendix:terms}. Listed in this Appendix are the full queries that generated papers that satisfied our inclusion criteria, listed below:
\begin{itemize}
    \item The papers are either published peer-reviewed works or reports published by non-profits, think tanks, or firms. 
    \item The papers clearly discuss the technical aspects of the AI model, which must include the specific model method (machine learning, generalized linear model, deep learning), relevant predictor(s), and outcome(s). 
    \item The papers are not literature reviews or surveys.
    \item The papers are written in English.
    \item The papers outline a model(s) with tangible implementation in a health system.
    \item The papers reference `cost' in a monetary, energy, time, or labour context. This does not include papers that solely perform cost prediction (cost as outcome variable), which need an additional consideration of cost burden.
\end{itemize} 

We chose to exclude survey papers due to their lack of technical specifications for models (due to their focus on higher-level trends across model performance), and their incorporation/comparison of models with distinct goals. While certain survey papers may have satisfied our remaining criteria, the models analyzed within them often did not clearly discuss technical AI design aspects or were not explicitly relevant to a health system goal. All papers in our database considered at least one dimension of cost, as outlined in Table \ref{table:2}, with papers being multi-counted as necessary.
There were no time constraints used in our search. We searched for papers across the ACM Digital Library, IEEE Xplore, Springer, JMIR, PubMed, and Elsevier to collect articles from across disciplines and health system locations. From the initial list of papers we also examined reference citations to identify any relevant literature (also following the above criteria).  In total, we assembled 114 peer-reviewed works and reports following our criteria, of which most were situated in the US (n=42), India (n=14), China (n=13), UK (n=8), and Canada (n=3), among other locations. Our corpus breakdown by database can be found in Appendix \ref{appendix:terms}.

Those papers outlining a model but without defined predictors, method, or target outcome were removed. Using the PRISMA statement \cite{prisma09,prisma20}, we iterated through relevant literature by title, abstract, and publication content. This brought our initial list of 450 papers down to 114 (flow diagram in Appendix \ref{sec:appendix_graph}).

\begin{table}[t]
\begin{tabular}{llll}
\multicolumn{3}{l}{\textbf{Classification}}                   & \textbf{n} \\ \hline
\multicolumn{3}{l}{\textbf{Peer Reviewed}}                    & \textbf{112}         \\
\textbf{}       & \multicolumn{2}{l}{\textbf{Journal}}                 & \textbf{79}         \\
                &             & Digital medicine              & 37         \\
                &             & CS                            & 19          \\
                &             & Health informatics            & 14          \\
                &             & Health                        & 7          \\
                &             & Economics                     & 2          \\
\textbf{}       & \multicolumn{2}{l}{\textbf{Conference}}              & \textbf{33}         \\
                &             & CS                            & 28         \\
                &             & Health informatics            & 2          \\
                 &             & Health             & 1          \\
                &             & Engineering                   & 1          \\
                &             & Business             & 1          \\
\multicolumn{3}{l}{\textbf{White Papers}}                     & 2          \\
                &             & Health                        & 2          \\ \hline
\multicolumn{3}{l}{Total}                                     & 114                             
\end{tabular}
\caption{Descriptive Dataset Characteristics}
\label{table:1}
\end{table}

\subsection{Data Analysis} \label{sec4.2}

The first and last authors narrowed down the initial collection of papers to our corpus of 114, as per the approach in Appendix \ref{sec:appendix_graph}. After this, the first author read all the papers and conducted primary analysis using iterative thematic analysis \cite{Braun_grounded,Clarke2017}. Multiple rounds of conversation were conducted with the last author to define requirements for the recruited papers and determine appropriate codes and dimensions. During such conversations, papers were coded according to a preliminary set of dimensions outlined in Table \ref{table:2}, with any confusions resolved in conversation. Papers were initially selected based on title and abstract, then read fully to ensure a health setting and consideration of cost. The selected papers were then read with specific attention paid to the methods, input data, and outcomes, expanding Table \ref{table:2} to include more dimensions agreed upon by all co-authors. The papers were also categorized by venue and field (Table \ref{table:1}). As more papers were collected, further intricate dimensions were defined in Table \ref{table:2} (distinguishing ``Reduce Healthcare Spending'' and ``Cost-Effectiveness''; ``Demographics'' and ``EHR''). Those papers using multiple methods, predictors, outcomes, or cost considerations were coded under multiple dimensions throughout our tabulations. We similarly constructed our discussion points, iteratively extracting and developing conclusions around model costs, sustainability concerns, and human-centered principles. Each of the models' predictors, methods, and outcomes were closely examined and coded as per Table \ref{table:2}, to succinctly describe the (often overlapping) technical components. During our iterative co-author discussions, any ambiguities for appropriate coding were resolved, ahead of data analysis and the construction of results. This prepared an agreed-upon corpus, set of codes, and coded corpus with which to execute our analysis on. Our PRISMA approach is outlined in Appendix \ref{sec:appendix_graph}, with cross-tabulation tables in Appendix \ref{sec:tables_appendix}. Out of the 114 papers, 57 (50\%) examine only one AI model, while the other 57 (50\%) test multiple models for the same outcome, with identical or similar data, to examine the optimal methodology. 

The codes identified early in the process were expanded to include method dimensions such as image processing, once a suitable body of computer vision-related work was identified. The list of predictors was repeatedly refined as many categories overlap (EHR, demographics, hospital admissions) but are not always used together. No distinction was made between age, gender, and race predictors since models used them in combination together under demographic information. Inferential statistics encompasses those models with a generalized linear model (GLM) or statistical test (ST) basis, while machine learning (ML) includes models such as support vector machines, ensemble boosted methods, or decision tree-related models. 

Cost considerations were coded alongside the other dimensions. Those papers acknowledging implementation of the model as a reduction of healthcare spending were coded as such, as were those models presented as cost-effective alternatives, and those recognizing modeling-related costs.

We also investigated if authors utilized HCAD principles \cite{Baumer17}, a popular strategy for recent SIGCHI literature reviews for public interest fields \cite{Moon2024,mcconvey2023,saxena_lit,pam_review_1,pam_review_2,huber_public}. 
We concur with their conclusion that only a small portion of models actively consider theoretical, speculative, and participatory approaches in development, encouraging designers and developers to pursue human-centered modeling (Table \ref{table:2}). 


\begin{table}[t]
\centering
\resizebox{\textwidth}{!}{\begin{tabular}{l|l|l|l|l}                                  & \textbf{Dimension}                   & \textbf{Code}                     & \textbf{Count} & \textbf{\%} \\ \hline
\textbf{Computational Method} & \textbf{Machine Learning (ML)}       & Supervised Learning               & 99             & 86.8        \\
                              &                                      & Unsupervised Learning             & 15              & 13.2         \\
                              &                                      & Natural Language Processing (NLP) & 8              & 7.0         \\
                              & \textbf{Inferential Statistics (IS)} & Generalized Linear Models (GLM)   & 36             & 31.6        \\
                              &                                      & Statistical Tests                 & 22             & 19.3        \\
                              & \textbf{Deep Learning (DL)}          & Neural Network                    & 62             & 54.4        \\
                              & \textbf{Image Processing}            & Computer Vision                   & 10              & 8.8        \\ \hline
\textbf{Predictor Variables}  & \textbf{Demographics}                & Age, gender, race                 & 73             & 64.0        \\
                              & \textbf{Hospital Admissions}         &                                   & 41             & 36.0        \\
                              & \textbf{EHR}                         &                                   & 38             & 33.3        \\
                              & \textbf{Laboratory Results}          &                                   & 28             & 24.6        \\
                              & \textbf{Claims Data}                 & Insurance Claims                  & 22             & 19.3       \\
                              & \textbf{Physician Notes}             & Unstructured                      & 16             & 14.0        \\
                              & \textbf{Prescriptions}               &                                   & 17             & 14.9        \\
                              & \textbf{Images}                      & Diagnostic Imaging                & 17              & 14.9        \\ \hline
\textbf{Outcome Variables}    & \textbf{Outcome}                     & Condition Diagnosis               & 37             & 32.5        \\
                              &                                      & Patient Feature Analysis        & 28             & 24.6        \\
                              &                                      & Risk Prediction                   & 15             & 13.2        \\
                              &                                      & Cost Prediction                   & 11              & 9.6        \\
                              &                                      & Patient Experience                & 3              & 2.6         \\ \hline
\textbf{Cost}                 & \textbf{Cost Considerations}         & Reducing Healthcare Spending      & 73             & 64.0        \\
                              &                                      & Cost-Effectiveness                & 59             & 51.8        \\
                              &                                      & Model Cost                        & 39             & 34.2 \\
                              \hline
\textbf{Human-Centered}                 & \textbf{Design Principles \cite{Baumer17}}         & Theoretical      & 25             & 21.9        \\
&                                      & Speculative                & 10             & 8.8        \\
&                                      & Participatory                        & 7             & 6.1       

\end{tabular}}
\caption{Technical, Cost, and Human-Centered Codebook. Percentages represent the Count divided by the total 114 papers in the corpus.}
\label{table:2}
\end{table}

\section{Results}

We begin by outlining the results of our literature review with a summary of the descriptive characteristics of our dataset, then we describe our results for each research question. We cross-tabulate between model predictors, methods, outcomes, and cost considerations to examine the relationships between each of these components of algorithm design in health systems.

\subsection{Dataset Characteristics} \label{sec5.1}

Our corpus of 114 papers includes 112 peer-reviewed publications and 2 reports (Table \ref{table:1}). 57 papers (50\%) examine a single AI model, while the other 57 (50\%) compare multiple models for the same task (ex. comparing neural network with random forest to detect coronary heart disease \cite{ACM_Banerjee}). The highest number of models compared in a single paper is 16: 15 machine learning classifiers and one neural network, compared in terms of accuracy and specificity to predict a COVID-19 test result \cite{Springer_Kamari}. Most of the papers come from computer science (n=47), digital medicine (n=37), and health informatics (n=16). Digital medicine refers to the usage of digital solutions and computational processes (AI, ML) for health systems \cite{digital_medicine}. 49 papers were developed for health system support, 41 for physician support, and 12 for direct patient support. When examining which papers incorporate cost concerns, 73 convey how AI can reduce healthcare costs, 59 consider AI to be cost-effective solutions, and 39 explicitly consider model costs (Table \ref{table:2}, with papers being multi-counted across these dimensions). 
The two reports are published by McKinsey \& Company \cite{McKinsey} and the RAND Corporation \cite{RAND_Heins}. Of the 114 papers, 4 (3.5\%) are reported to be actively implemented in the relevant health system, with others encountering barriers such as the need for extensive validation \cite{Nature_Liu,Elsevier_Lu,Elsevier_Salmons_1,Elsevier_Salmons_2,PubMed_Smak_Gregoor}, further AI explainability \cite{PubMed_Soliman,ACM_Zhang_Explain,Elsevier_Ismukhamedova}, ethical considerations \cite{Elsevier_Sanderson}, and cost evaluations \cite{PubMed_Feretzakis,JMIR_Rajkumar,ACM_Zhang_Explain}.

\begin{figure}[t]
\includegraphics[scale=0.6]{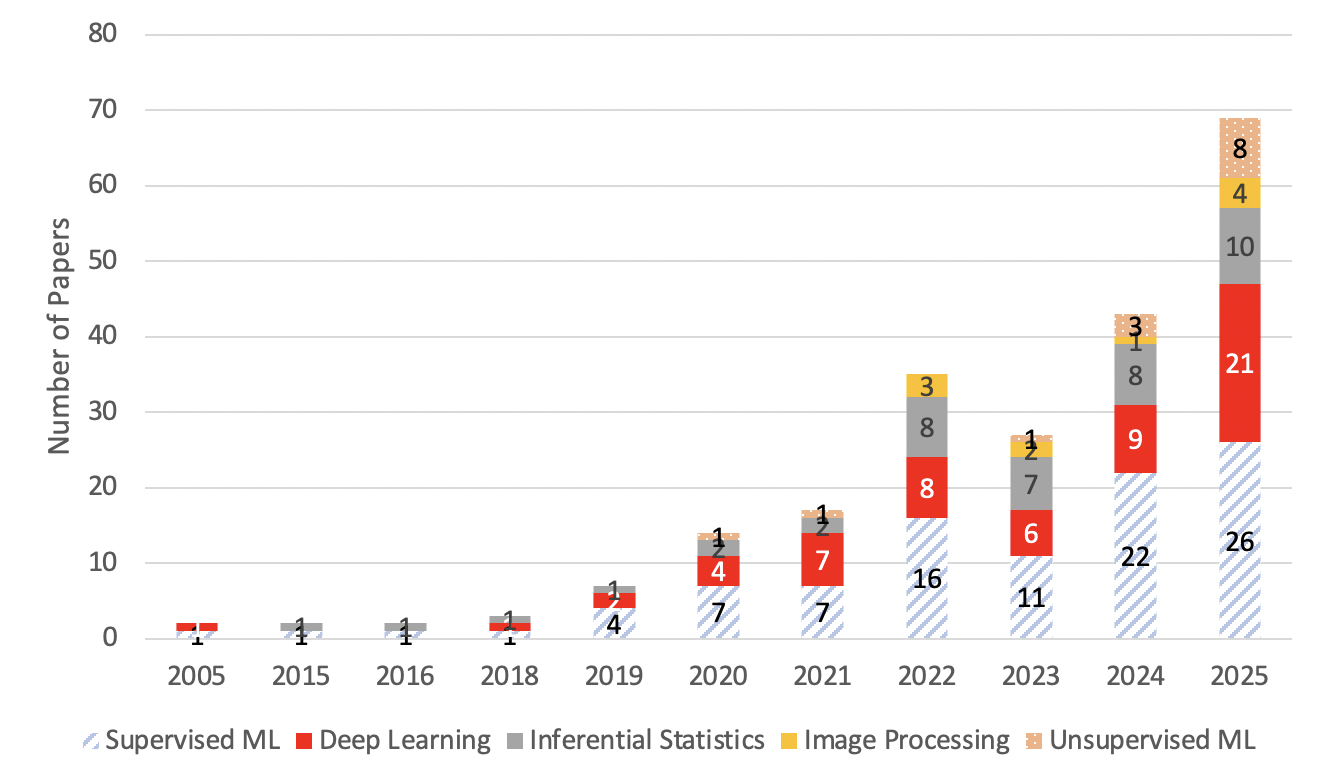}
\caption{Methods Used to Build Models}
\label{fig:methods}
\end{figure}

\subsection{Technical Model Design} \label{sec5.2}

\subsubsection{Methods Deployed in Models for Health Systems} \label{secmethods}

The models in our papers deploy three main methods: \textbf{supervised ML, inferential statistics, and deep learning} (neural networks). Inferential statistics (IS) includes those models based on statistical tests or generalized linear models (GLMs), while ML includes techniques such as random forest, gradient boosting, and support vector machines. Models incorporating neural networks are classified as deep learning (DL). 
Those papers using multiple methods were counted under both/all codes (ex. developing a decision tree and GLM in a single paper would be counted as both supervised ML and IS in Figure \ref{fig:methods}, Table \ref{table:2}, and Appendix \ref{sec:tables_appendix}).

\paragraph{Inferential Statistics}

A total of 43 papers use IS methods. 15 of these papers explore statistical tests (STs) and GLMs in combination, while seven consider only STs and 21 consider only GLMs. IS methods have been consistently used throughout the past decade, with applications in health system, clinician, and patient support. These models assist condition diagnosis (n=13), patient feature analysis (n=12), cost prediction (n=8), risk prediction (n=4), and patient experience (n=2). The most frequent data inputs for these models are demographic (age, sex, ethnicity), EHR, and hospital admission information. 

\paragraph{Machine Learning and Deep Learning}

Throughout the study timeline, supervised ML methods were used steadily across years, while DL and unsupervised ML have increased in proportion from 2018 to 2025. 
Some of these unsupervised methods include clustering of heart failure patients \cite{ACM_Banerjee}, and large language models (LLMs) to examine patient experiences with chatbots \cite{ACM_Jo,IEEE_Singh}. 
Random forest (n=40), eXtreme Gradient Boosting (XGBoost) (n=22), and decision trees (n=17) were the most popular ML methods. Only one paper considered random forest modeling alone \cite{Elsevier_Haeberle_Koch}, while the other 39 compared random forests to GLMs, decision trees, and neural networks. Similarly, only one paper considered decision tree modeling alone \cite{PubMed_Soliman}, while the other 16 included comparisons with neural networks and logistic regression. Five papers explored XGBoost alone \cite{ACM_Gyldenkaerne,IEEE_Gupta,Elsevier_Zea-Vera,Lin_Bai_Huang_Lee_Vu_Chiu_2025,Xiao_Li_Wang_Wang_Chen_2025}, while the other 17 compared this method with k-nearest neighbors and logistic regression. 

\paragraph{Image Processing} \label{secip}

The advancement of computational power to analyze unstructured data has made image processing (IP) more prevalent in healthcare, for use in settings such as diabetic retinopathy screening \cite{xie_dr}, digital pathology \cite{mota_cost}, and radiological examinations \cite{mota_cost}. Since IP is prevalent across our identified method categories (IS, supervised ML, neural networks), we consider IP as a unique code, spanning all previously defined groups, noting 
a recent interest in IP via neural networks \cite{image_proc_book,image_nn,image_nn_2}. Outside of ML, IP has been used to note important features, remove noise, and filter relevant images \cite{image_proc_not_ML,image_proc_not_ML_2}. Image processing entered this study corpus in 2020, using diagnostic imaging data (CT, X-ray, ultrasound images) in seven papers \cite{Bharati_R_Singh_Khanna_V_C_2025,Lyth_Gialias_Husberg_Bernfort_Bjerner_Wiberg_Levin_Gustafsson_2026,ACM_Barillaro,IEEE_MS_Thisin,IEEE_Sinha,PradeepKumar_J_B_K_2025,Siam_Ahmed_Khan_Islam_Milon_Ahamed_Islam_2025}, skin lesion images in two papers \cite{IEEE_Ezenkwu,PubMed_Smak_Gregoor}, and tooth images in the remaining paper \cite{PubMed_Schwendicke}. IP highlights an additional cost concern of image quality and the need to repeat images due to unreadable outputs, which can double or triple image acquisition costs per patient \cite{xie_dr}.

\subsubsection{Predictors Deployed in Models for Health Systems} \label{secpredictors}

\begin{figure}[t]
\includegraphics[scale=0.55]{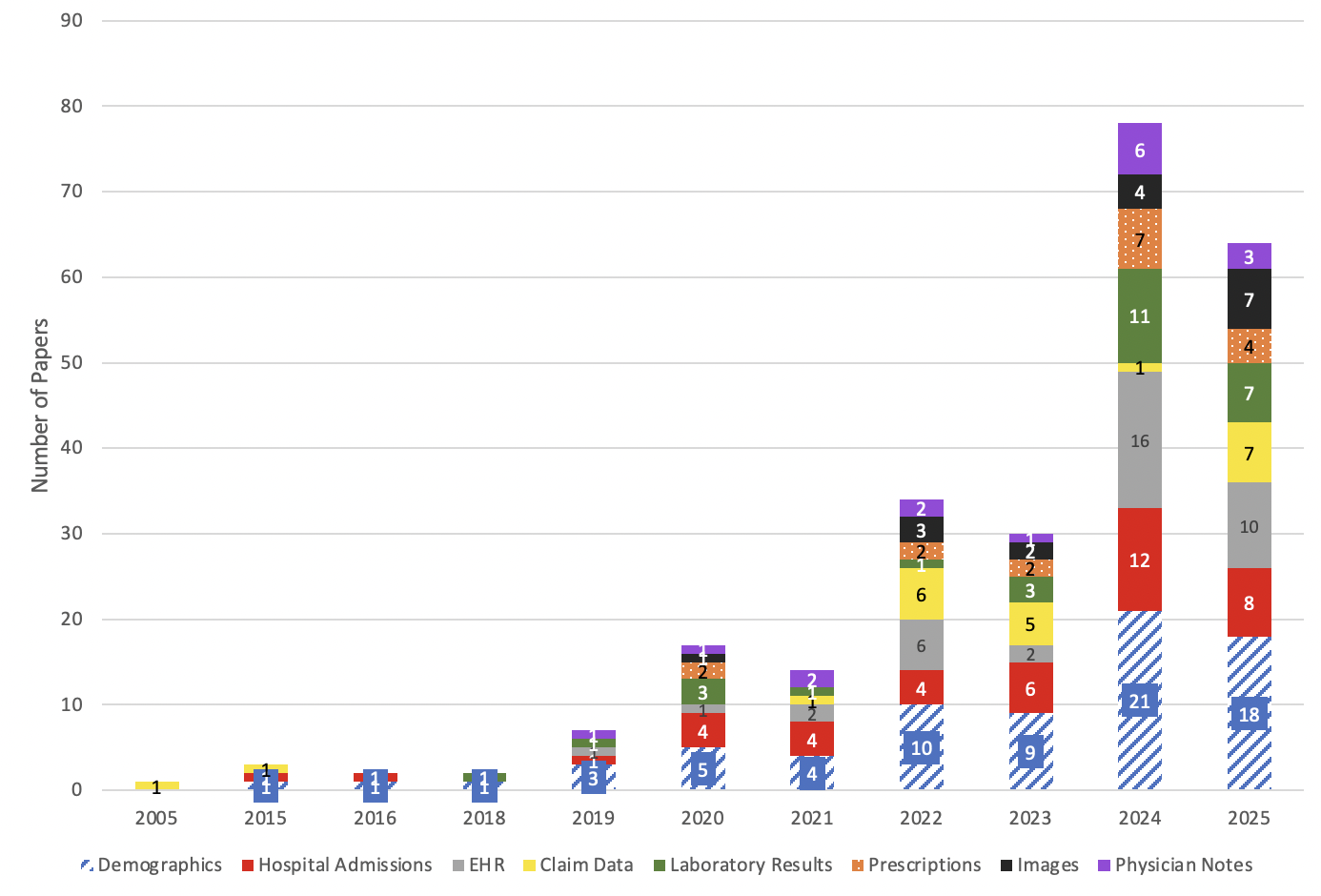}
\caption{Predictors Used to Build Models}
\label{fig:predictors}
\end{figure}

The wide range of predictors in our corpus are coded in eight dimensions: \textbf{EHR, Claims Data, Images, Demographics, Physician Notes, Prescriptions, Laboratory Results,} and \textbf{Hospital Admissions}. While there is often overlap between categories (most frequently between EHR, demographics, and hospital admissions), many papers only use a single code.

Demographic (age, sex, race, n=73, 64.0\%) and hospital admission (n=41, 36.0\%) information have been used in almost every year of the study timeline, often collected in electronic format. Insurance claim data (n=22, 19.3\%) and EHRs (n=38, 33.3\%) are being increasingly used, frequently to understand how costs of care visits change due to other factors. Images (n=17, 14.9\%) are being used for condition diagnosis and classification. The spread of predictors has increased, particularly since 2021 (Figure \ref{fig:predictors}), as models use combinations of predictors to complement method choices. Those papers utilizing multiple predictors (nearly all papers in our corpus) were coded and counted under all applicable codes in Figure \ref{fig:predictors} and Appendix \ref{sec:tables_appendix}.
Creating data narratives from unstructured data is becoming a more popular strategy to incorporate AI into predictive tasks \cite{Williams_NLP_unstructured, Sedlakova_unstructured_narratives}. With the emergence of computational methods for unstructured data collection and analysis, physician notes (n=16, 14.0\%) regarding a patient's support system, medication history, or lifestyle can be synthesized for advanced patient personalization and storytelling potential \cite{Williams_NLP_unstructured,personalized_AI_costs_Ahmed}. 

\begin{figure}[t]
\includegraphics[scale=0.57]{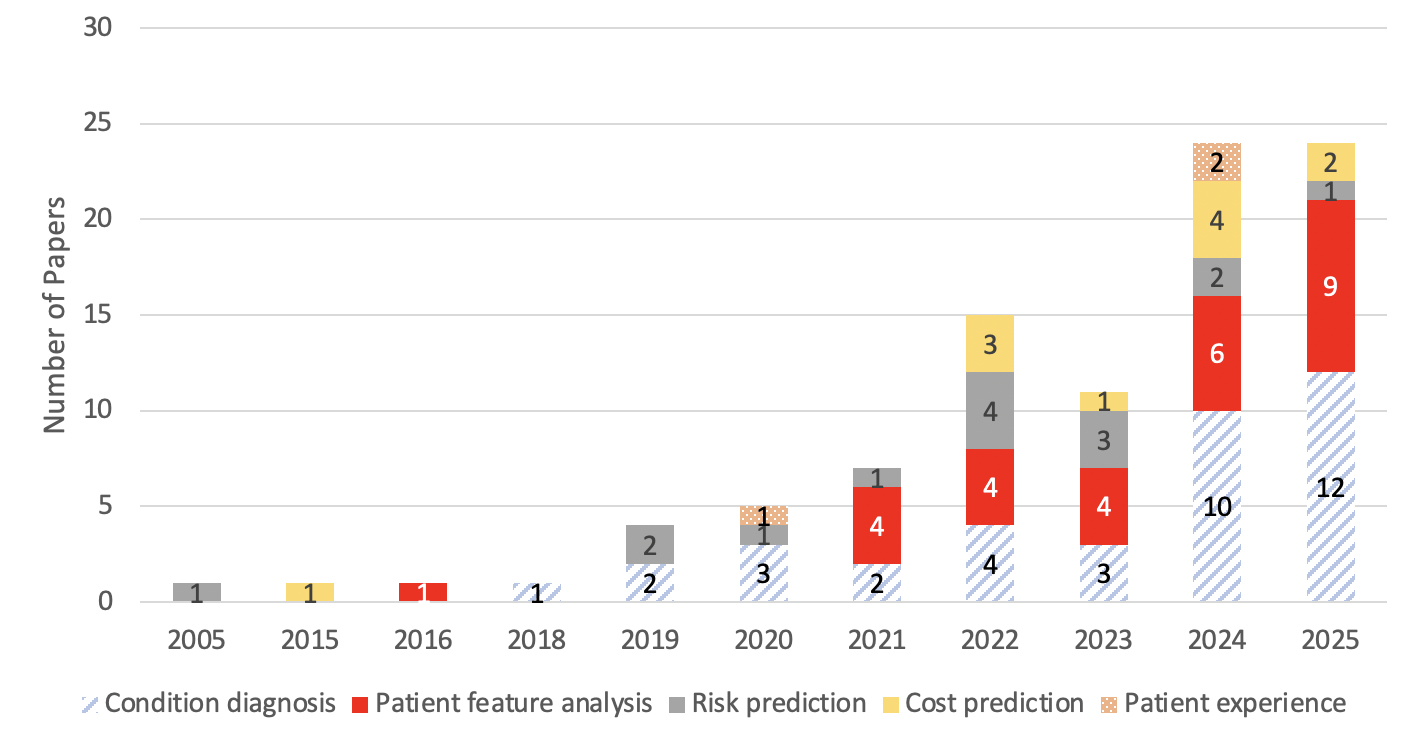}
\caption{Outcome Variables for the Models}
\label{fig:outcomes}
\end{figure}

\subsubsection{Outcome Variables in Models for Health Systems} \label{secoutcome}

From 2018 to 2025, AI models were consistently used for \textbf{condition diagnosis} (37, 32.5\%), \textbf{patient feature analysis} (28, 24.6\%), and \textbf{risk prediction} (15, 13.2\%). Risk prediction is a common outcome from 2019 onwards, with tasks including readmission risk (n=4, 3.5\%), disease risk (n=9, 7.9\%), or risk of certain patient behaviours (no-show, hospital stay length, n=2, 1.8\%) \cite{ACM_Hon,Elsevier_Mohanty,Elsevier_Meng,JMIR_Mens,IEEE_Pius,Nature_Zhao,JMIR_Oates}. Models that analyze \textbf{patient features} (frailty, level of need, length of hospital stay, comorbidities, mismedication, n=28, 24.6\%) are increasingly used from 2021 onwards, potentially due to the advent of blockchain methods and EHR efficiency (Figure \ref{fig:outcomes}) \cite{IEEE_Gupta,McKinsey,Elsevier_Mohanty}. These papers aim to reduce costs by more clearly understanding the patient and analyzing their predicted behaviour to allocate staff and resources. \textbf{Patient experience analyses} (examining how patients interact with patient portals \cite{ACM_AlRamahi} and chatbots \cite{ACM_Jo,IEEE_Singh}) investigate why patient portals and chatbots are underused, and which features can make them more accessible and popular with multiple patient groups. 
The ability to combine data sources has allowed for an increase in personalized medicine efforts, contributing to a body of work on how to apply different computational methods to personalized care
\cite{AI_Personalized_Gifari,personalized_AI_Johnson,personalized_AI_costs_Ahmed}.

\subsection{Cross-Tabulation of Predictors, Methods, and Outcomes}

\subsubsection{Methods and Predictors}

The most frequent method in our corpus is supervised ML, most often utilizing random forest (40 papers, 35.1\%) and XGBoost modeling (22 papers, 19.3\%). All but four publications utilizing random forest modeling use demographic information as a predictor, alongside common combinations between EHRs, clinical notes, claims data, and hospital admissions. Similarly, all except four XGBoost publications use demographic input features, and frequently combined this with hospital admissions information (Appendix \ref{sec:tables_appendix}). 

Demographic information (age, sex, ethnicity) is the most common predictor for all modeling except IP, seen in 88.4\% of IS, 63.9\% of ML, and 51.6\% of DL publications. Two of the ten IP studies include demographic information alongside the diagnostic (or other) imaging used for predictions \cite{IEEE_MS_Thisin,PubMed_Schwendicke}. This information is commonly collected from patients during interactions with the health system, which makes it widely accessible, but comes with its own risks such as inconsistent reporting or sharing of this information by developers \cite{bozkurt_demographics}. Hospital admissions data (such as date and time of admission or hospital stay length) is the next most frequent predictor across all methods, often used alongside EHR data. Admissions data is imperative for studies examining patient care characteristics such as cost or demand of patient care \cite{Nature_Liu,Elsevier_Meng,IEEE_Pius,Elsevier_Zea-Vera}, readmission risk \cite{ACM_Hon}, and stay length \cite{IEEE_MS_Thisin}.

Proportionally, DL uses images as predictors more often than other ML methods, aligning with the expectation of AI to read images and make conclusions more efficiently and potentially more accurately than physicians \cite{naylor_deep}. Similarly, physician notes are used proportionately more frequently by DL methods than ML, to extract unstructured data input \cite{wang_deep}.

\subsubsection{Outcomes and Predictors}

Condition diagnosis is the most common outcome (37 papers, 32.5\%). These models are used to detect whether a patient had or has a particular condition, classified in binary or multi-class ways. Examples include coronary artery disease \cite{ACM_Banerjee}, preeclampsia \cite{Elsevier_Haeberle_Koch}, pneumonia in children \cite{Elsevier_Kanwal}, and COVID-19 \cite{Springer_Kamari,IEEE_Mistry}. Condition diagnosis predictors were most commonly demographic and laboratory information (Appendix \ref{sec:tables_appendix}). This complements the usage of AI models as cost-effective diagnosis tools, using data inputs that are easily accessible by clinicians. Eight of the ten IP papers in our corpus are coded under condition diagnosis.

Patient feature analysis (n=28, 24.6\% of papers) and risk prediction (n=15, 13.2\% of papers) are also frequent outcomes. In order to predict risk of a certain disease state or outcome, developers often utilize demographic, hospital admissions, and EHR data. Less frequently, claims data is used to predict risk of sepsis \cite{JMIR_Rogers}, diabetes \cite{ACM_Kudyba}, and cardiovascular disease \cite{ACM_Pfohl}. For patient feature analysis, prescriptions are used more frequently than for other outcomes, predicting readmission risk \cite{Elsevier_Mohanty} and patient outcomes \cite{Elsevier_Meng,Elsevier_Wei,Elsevier_Zea-Vera}.

Cost prediction papers (n=11, 9.6\% of papers) also use demographic, hospital admission, and EHR information as predictors, alongside lesser claims data input. No cost prediction papers utilize physician notes, laboratory results, or images as predictors (Appendix \ref{sec:tables_appendix}). Lastly, patient experience analyses use unique inputs in each study, consisting of chatbot responses \cite{ACM_Jo,IEEE_Singh} and patient portal reviews \cite{ACM_AlRamahi}, which were not coded in our study. 

\subsubsection{Methods and Outcomes}

Supervised ML stood alone as the most popular method choice in our study corpus. 
Unsupervised ML is used less frequently (Appendix \ref{sec:tables_appendix}) as are natural language processing (NLP) methods. 
Neural networks are the second most common method with a total of 62 papers (54.4\%), exploring all outcomes (Appendix \ref{sec:tables_appendix}). Most frequently, neural networks are used for condition diagnosis (n=28, 24.6\% of all papers) and patient feature analysis (n=12, 10.5\%). GLMs are used consistently across all outcomes and were one of the most widespread techniques in our corpus. 
Overall, 37 papers explore condition diagnosis (32.5\%). 35 of these papers use supervised ML, 27 of them in combination with DL. 28 papers explore patient feature analysis (19.7\%), split across GLMs, supervised ML, and DL. 15 papers involve risk prediction (13.2\%), which all use supervised ML. 11 explore cost prediction (9.6\%) and 3 examine patient experience (2.6\%).

\begin{figure}[t]
\includegraphics[scale=0.57]{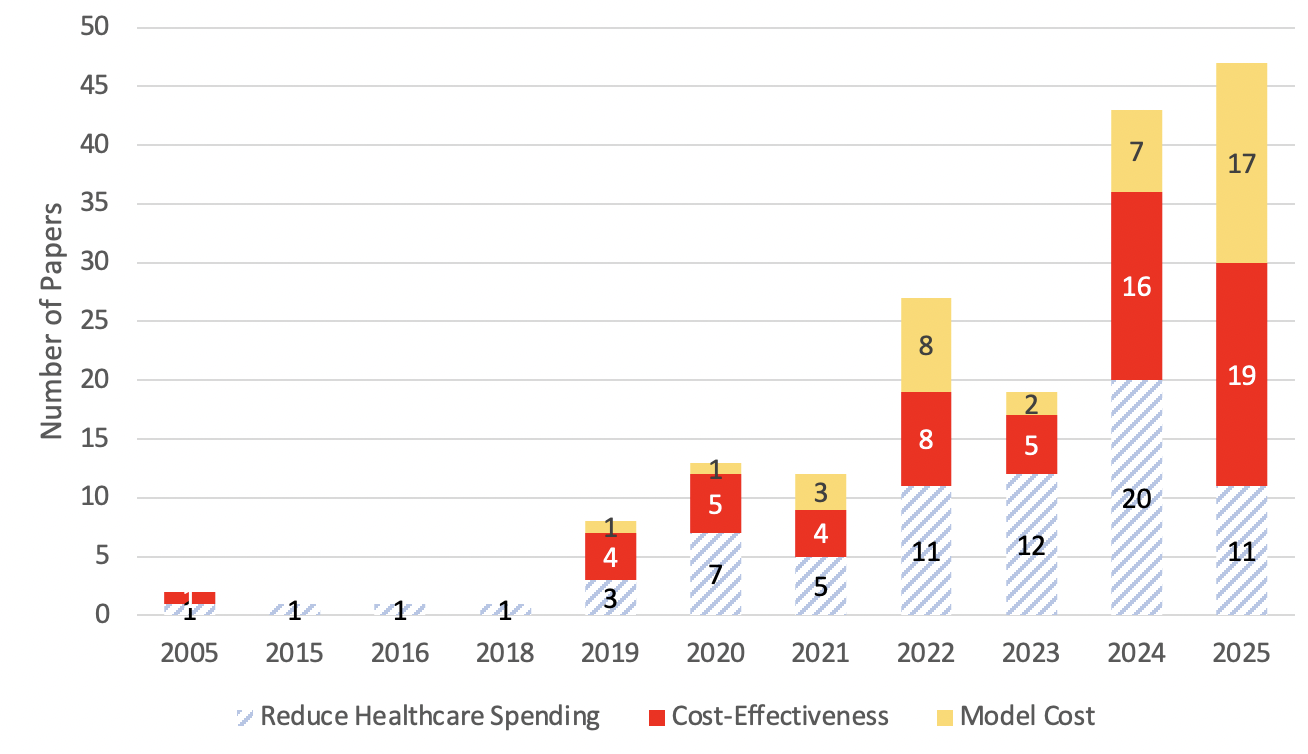}
\caption{Cost Considerations of the Models}
\label{fig:costs}
\end{figure}

\subsection{Cost Considerations in AI Models for Health Systems} \label{seccosts}

AI models have consistently been promoted to reduce healthcare spending and increase the efficiency of health system administrations and daily care practices (Figure \ref{fig:costs}) \cite{Kejriwal_cost}. From 2019, the models in our corpus have been used as cost-effective solutions for risk prediction, condition diagnosis, and prognosis, to assist healthcare decision-making in high-stakes environments \cite{Kejriwal_cost}. Simultaneously, AI-associated cost concerns have emerged, most frequently in the form of wider model costs, examining the cost of data collection and the electronic and personal time costs of incorporating these models into healthcare settings \cite{IEEE_Lee}. While AI has been explored and developed in order to reduce resource, staff, and care allocation efforts throughout the study timeline, cost-effectiveness and model cost discussions are becoming more pertinent (Figure \ref{fig:costs}). 

\subsubsection{Intersection of Model Costs and Model Design} \label{seccost_design}

We also review the different cost considerations made for each category of methods and outcomes. From Table \ref{table:5}, the most frequent cost-related motivation is to reduce healthcare spending (n=73, 64.0\%), prevalent for all methods and outcomes. Developers will present their work as an alternative to worker labour or as an effort to increase efficiency, through reduction of current costs to their health system. Costs are most frequently attributed to reducing expenditures such as patient care \cite{ACM_AlRamahi,ACM_Gyldenkaerne,ACM_HuangSong,ACM_Kudyba,ACM_Singh,JMIR_Mens}, readmissions \cite{ACM_Hon,ACM_Oota,JMIR_Rajkumar}, monitoring \cite{ACM_Banerjee}, and testing and diagnosis \cite{ACM_Bartenschlager,ACM_Mahmud}. Cost-effectiveness (n=59, 51.8\%) is considered for all methods and outcomes; however, this is not as prevalent of a concern as reducing healthcare spending. Particularly for cost prediction and GLM papers, there is less interest in reviewing the cost-efficiency lent by the model (Table \ref{table:5}). Lastly, model costs (n=39, 34.2\%) are least frequently considered across methods and outcomes. When model costs are considered they include consideration of running models \cite{ACM_Oota,IEEE_Gupta,IEEE_Mistry,ACM_Barillaro,Elsevier_Mohanty,JMIR_Rajkumar,IEEE_Ezenkwu,IEEE_Massa,Tung_Le_Yao_Huang_Lim_Sng_Lau_Tan_Chen_Tay_et_2025}, storage of data \cite{IEEE_Gupta,IEEE_Mistry,IEEE_Ezenkwu,JMIR_Rajkumar,Vusumuzi_Godwin_2025}, hardware and maintenance \cite{Elsevier_Wei,Do_Allison_Nguyen_Phung_Tran_Le_Nguyen_2024,IEEE_Ezenkwu,IEEE_Sinha,Elsevier_Mohanty}, and training models \cite{Elsevier_Sanderson,ACM_Lv,Elsevier_Li,Elsevier_Mohanty,PubMed_Schwendicke,Elsevier_Kanwal,IEEE_Massa,Wang_Chen_Wu_Jiang_Lin_Qiu_2025,Vusumuzi_Godwin_2025}. There is only one cost prediction paper that reviews, considers, or evaluates model costs (computational, hardware/software, or data-related) \cite{Zhang_Zhu_Chen_Wang_2025}.

\begin{table}[t]
\resizebox{\textwidth}{!}{\begin{tabular}{l|ccc|ccccc}
                            & \multicolumn{3}{l|}{{\textbf{Computational Method}}} & \multicolumn{5}{l}{\textbf{Outcome}}                                                                                                                                                                                                                                                                                                                               \\ \cline{2-9} 
\textbf{Cost Consideration} & \textbf{GLM}     & \textbf{ML}    & \textbf{DL}    & \textbf{\begin{tabular}[c]{@{}c@{}}Condition \\ Diagnosis\end{tabular}} & \textbf{\begin{tabular}[c]{@{}c@{}}Patient \\ Characteristic\end{tabular}} & \textbf{\begin{tabular}[c]{@{}c@{}}Risk \\ Prediction\end{tabular}} & \textbf{\begin{tabular}[c]{@{}c@{}}Cost \\ Prediction\end{tabular}} & \textbf{\begin{tabular}[c]{@{}c@{}}Patient \\ Experience\end{tabular}} \\ \hline
\textbf{Reduce Spending}    & 25               & 68             & 25             & 19                                                                      & 20                                                                         & 12                                                                  & 9                                                                   & 2                                                                      \\
\textbf{Cost-Effectiveness} & 16               & 57             & 34             & 26                                                                      & 13                                                                         & 7                                                                   & 2                                                                   & 3                                                                      \\
\textbf{Model Cost}         & 12               & 39             & 23             & 15                                                                      & 5                                                                          & 6                                                                   & 1                                                                   & 0       
\end{tabular}}
\caption{Cross-Tabulation between Outcome, Method, and Cost Considerations}
\label{table:5}
\end{table}

As seen in Table \ref{table:5}, there is a larger number of papers promoting a reduction in healthcare spending than those actively considering model costs, across all computational methods and outcomes. When researchers are emphasizing a reduction in expenditures, they are less likely to also investigate how costly the development, deployment, and maintenance of these models are to stakeholders. Of the 39 papers in our corpus that consider model costs, only 12 (30.8\%) of them also promote a reduction in healthcare spending, and only 20 (51.3\%) of them advertise `cost-effectiveness.' We further categorize these (overlapping) model costs into computational, organizational, and human/social (Table \ref{tab:costxHCAD}), where papers mentioning multiple cost categories are multi-counted. Computational costs include those related to model training, computational demand, data management, and more (as explored in \cite{Wang_Chen_Wu_Jiang_Lin_Qiu_2025,IEEE_Massa,PubMed_Schwendicke,Elsevier_Kanwal,Wang_Yang_Liang_Lee_Cao_2024,Vusumuzi_Godwin_2025}). Organizational costs refer to those arising from implementation and workplace changes (ex. \cite{Stephan_Hanselmann_Bajramovic_Schosser_Laxy_2025,Wu_Hu_Yan_Li_Li_Chen_Lin_Zeng_Li_Mo_et_2025,Vusumuzi_Godwin_2025}). Lastly, human and social costs affect individuals (healthcare providers and patients) and their relationships, as well as subsequent equity concerns (ex. \cite{Siam_Ahmed_Khan_Islam_Milon_Ahamed_Islam_2025,Vusumuzi_Godwin_2025}). Ideally, to maintain a fully comprehensive evaluation of AI cost impact, studies would meet all three criteria, and transparently evaluate their sustainability and/or fairness trade-offs, and consider to whom the costs will be incurred (whether administrators, patients, taxpayers, or other groups).

\section{Discussion}

\subsection{Increased computational power strains model expectations and outcome tensions} \label{sec6.1}

Growing computational potential through large datasets and greater computational power will allow for increased AI applications in health systems, utilizing vast numbers of predictor, method, and outcome combinations \cite{mota_cost}. This expansion in modeling will likely make AI more easily available across health systems; however, it may continue to introduce biased recommendations and outcomes, sustaining previous tensions between stakeholder expectations and actionable insights.

As seen in Figure \ref{fig:predictors}, the diversity of predictors used in AI modeling has steadily increased since 2021. 
As a result of prioritizing data-driven AI, thus using all available data and vast computational power for more complex modeling (\ref{secpredictors}), we anticipate model costs to also increase \cite{data_centric}.
With greater amounts of modeling in health systems, other challenges such as bias and justice concerns continue to arise (\ref{seccost_design}) \cite{obermeyer_bias,Dankwa_justice}. These models may not have been considered before, as developers were restricted by available or processable data formats within existing computing machinery. AI has been presented as an ``inevitable evolution'' for health systems \cite{cadamuro_inevitable}, as previous barriers (unstructured or image input data, obtaining large datasets) are no longer constraints (\ref{secip}). This evolution in health systems may expand current modeling ecosystems through novel combinations of predictors, methods, and outcomes. 
There should also be the consideration of tensions existing between and within stakeholder groups regarding the human values they prioritize. When model intentions and implementation strategies do not align with administrators' needs, models are less likely to be trusted and create a larger organizational tension for workers on the ground \cite{saxena_bureau}, ultimately creating discord between users and commissioners and decreasing intended model efficiency. In parallel, these downstream effects may disrupt patient care journeys, potentially leading to unethical treatment in the form of inaccurate risk assessments, ineffective treatment plans, or delayed decision-making \cite{obermeyer_bias,Dankwa_justice}.
The methodology, development, or promotional strategies of health system models that are ``equitable'' or ``fair'' to one group may conflict with another's perspective, leading to conflict that may not be resolved through compromise, as previously seen in child welfare \cite{saxena_risk}. 
With greater potential for unique AI models in health systems, using a plethora of input variables (\ref{secpredictors}) and modeling techniques (\ref{secmethods}), designers and developers need to collaborate on which combinations of model components will best fit stakeholder needs, and consider other outputs of these solutions, including undesired costs.



\subsection{Models that reduce healthcare spending rarely consider computation-related costs} \label{sec6.2}

\begin{table}[t]
\begin{tabular}{l|lll}
                        & \textbf{Theoretical (n=9)} & \textbf{Participatory (n=1)} & \textbf{Speculative (n=5)} \\ \hline
\textbf{Computational}  & 7                          & 1                            & 3                          \\
\textbf{Organizational} & 5                          & 0                            & 1                          \\
\textbf{Human/Social}   & 2                          & 0                            & 1                         
\end{tabular}
\caption{Count of Model Cost Categories by HCAD Principle}
\label{tab:costxHCAD}
\end{table}


While reduction of healthcare spending is mentioned in our corpus throughout the study timeline, and cost-effective models are presented from 2019 onwards at consistent rates, those models aware of computational or model costs are seen less frequently (Figure \ref{fig:costs}). \textbf{Our results show that only a handful of papers from 2019-2025 examine model costs, with researchers most aware of costs associated with running models} \cite{ACM_Oota,IEEE_Gupta,IEEE_Mistry,ACM_Barillaro,Elsevier_Mohanty,JMIR_Rajkumar,IEEE_Ezenkwu,IEEE_Massa,Tung_Le_Yao_Huang_Lim_Sng_Lau_Tan_Chen_Tay_et_2025}\textbf{, storing data} \cite{IEEE_Gupta,IEEE_Mistry,IEEE_Ezenkwu,JMIR_Rajkumar,Vusumuzi_Godwin_2025}\textbf{, hardware and maintenance} \cite{Elsevier_Wei,Do_Allison_Nguyen_Phung_Tran_Le_Nguyen_2024,IEEE_Ezenkwu,IEEE_Sinha,Elsevier_Mohanty}\textbf{, and training models} \cite{Elsevier_Sanderson,ACM_Lv,Elsevier_Li,Elsevier_Mohanty,PubMed_Schwendicke,Elsevier_Kanwal,IEEE_Massa,Vusumuzi_Godwin_2025,Wang_Chen_Wu_Jiang_Lin_Qiu_2025}. From our examination of model cost dimensions across outcomes (Table \ref{table:cost_breakdown}), their are varying levels of acknowledgment of these costs throughout our corpus, with computational components being considered at higher rates than organizational and human/social costs. We especially highlight the lack of model cost concerns from cost prediction studies (Table \ref{table:cost_breakdown}). Only 25 of the 99 papers (25.3\%) that emphasize a reduction in healthcare spending or cost-efficiency actually mention how model costs are allocated and accounted for (\ref{seccost_design}). From our corpus, we generated a list of modeling-related costs. When a monetary, computing, resource, or labour expenditure was referenced, we noted any provided definition or measurement, extracting higher-level trends into our four dimensions (Table \ref{tab:framework}). These guidelines serve to assist developers with providing a holistic assessment of model costs, as generated by the trends in our review. This table expands on Table \ref{tab:costxHCAD} with the Financial dimension, which was not explicitly present in the papers in our corpus, but is another prominent cost consideration. This breakdown of potential model cost components presents taxonomy, potential measurements, and prompting questions for AI designers in health, pushing for standardized reporting of these components throughout development. At each stage of the model lifecycle, we encourage consideration of these costs and transparent communication with stakeholders. Each dimension in Table \ref{tab:framework} may fall to a different role, with administrators potentially carrying the organizational burden and downstream workers facing the human/social cost, while the health system itself may compensate the computational and overarching financial costs.



\begin{table}[t]
\centering
\begin{tabularx}{\textwidth}{
    l
    >{\raggedright\arraybackslash}X
    >{\raggedright\arraybackslash}X
    >{\raggedright\arraybackslash}X
}
\toprule
\textbf{Dimension} &
\textbf{Definition} &
\textbf{Guideline} &
\textbf{Example Measurements} \\
\midrule
Financial &
Monetary expenditures &
Is financial cost reported? &
Development, licensing, maintenance \\

Computational &
Computing resources  &
Is computational cost reported? &
Training, GPU, data storage, energy use \\

Organizational &
Resources required for adoption &
Is organizational burden reported? &
Staff training, deployment time, implementation effort \\

Human/Social &
Time and labour required by users or providers &
Is human or social burden reported? &
Workflow burden, usability, staff time \\
\bottomrule
\end{tabularx}
\caption{Model Cost Dimensions and Example Measurements}
\label{tab:framework}
\end{table}


Additional trade-offs exist between model cost and equitable AI performance. Prioritizing human-centered design principles includes examining subgroup model performance, often through algorithmic audits \cite{audit1,audit2,audit3}. The cost-quality trade-off of a more expensive model performing more fairly across majority and minority groups is an ongoing avenue for HCAD research, without a clear monetary price associated with incremental subgroup performance. However, we encourage researchers to prioritize calibration across subgroups, especially for vulnerable populations frequently facing sociotechnical and structural barriers to healthcare access and resources \cite{barrier1, barrier2, barrier3, barrier4}. The most cost-effective AI model, requiring the fewest time and resources, may produce the least equitable results. Short-cutting model iteration, validation, testing, and detailed design steps, this model may be beneficial to health system administrators' workflows, whilst leading to dramatic, inequitable downstream effects. On a situational basis, designers should emphasize transparent evaluation of this trade-off, pursuing human-centered principles in manner that maintains an accessible and beneficial model.



With AI ecosystems growing past what we have previously considered the bounds of computational purpose in health systems, the increasing data and modeling volumes prompt reflection of the expanding electricity requirements, CO2 emissions, and resources allocated to automated processes  \cite{sustain_old_baumer,Van_Wynsberghe_sustainability}. Through the use of resource-intensive procedures and single-use devices, health systems generate large amounts of waste contributing massively to climate change \cite{waste}. AI modeling can reduce its impact on the carbon footprint of health systems by considering \emph{computational} and \emph{organizational} requirements (water, carbon emissions, electricity, hardware, time and labour for implementation), and minimizing excess modeling, agglomerating only necessary data, and reducing bias in initial training stages \cite{sustain_old_baumer}.
Beyond the clinical front lines lie additional but often invisible \emph{social} costs of AI production: the digital labour of data labelling, frequently outsourced to low‑wage workers in the Global South \cite{digital_labour_3}. When these costs remain off‑ledger, decision makers underestimate the true price of “efficiency,” much as developers do when they ignore externalized harms \cite{chadli}. van Wynsberghe [2021] urges this generation of AI to ``place sustainable development at its core,'' to realistically estimate modeling-related resource allocation across communities \cite{Van_Wynsberghe_sustainability}.  Combining human-centered strategies with a sustainable focus can decrease model costs, through raised awareness of relevant resource-intensive modeling efforts \cite{Van_Wynsberghe_sustainability}.  However, consideration must be paid to those models that are `higher-cost' (financially, computationally, organizationally, socially) but prioritize sustainability and potentially reduce environmental impact upon implementation. A model cannot be labeled `most optimal' due to a minimized financial burden, instead developers should refer to Table \ref{tab:framework} to explore and report a holistic cost framework for their AI design. While a model may claim to be a premier and cost-effective solution, downstream social impacts, workplace burdens, and accountability may dynamically surface through unintended consequences. 
AI solutions for vulnerable populations need to anticipate equitable outcomes, including both the quality of solutions  \textit{and} each dimension of cost benefits allocated between demographically-defined groups (Table \ref{tab:framework}).


\begin{table}[t]
\resizebox{\textwidth}{!}{
\begin{tabular}{l|rrrrr}
& \multicolumn{1}{l}{\textbf{Condition Diagnosis}} & \multicolumn{1}{l}{\textbf{Patient Characteristic}} & \multicolumn{1}{l}{\textbf{Risk Prediction}} & \multicolumn{1}{l}{\textbf{Cost Prediction}} & \multicolumn{1}{l}{\textbf{User Experience}} \\ \hline
\textbf{AUC-ROC}              & 27                                               & 14                                                  & 9                                            & 5                                            & 1                                            \\
\textbf{Accuracy}             & 21                                               & 7                                                   & 6                                            & 1                                            & 0                                            \\
\textbf{F1}                   & 12                                               & 8                                                   & 0                                            & 2                                            & 1                                            \\
\textbf{R\textsuperscript{2}} & 0                                                & 6                                                   & 3                                            & 4                                            & 4                                            \\
\textbf{MAE}                  & 0                                                & 3                                                   & 0                                            & 4                                            & 0   
\end{tabular}}
\caption{Model Evaluation Metrics by Outcome}
\label{table:metrics}
\end{table}

\subsection{Human-centered AI models question important computational context and downstream impact} \label{sec6.3}

Using human-centered algorithm design (HCAD) strategies \cite{Baumer17} throughout model development integrates stakeholder feedback, examines true needs, and defines achievable expectations for AI solutions. Human-centered strategies contribute to sociotechnically-aware models, incorporating themselves more easily into existing infrastructure \cite{Baumer17}. 
This engagement can be defined in terms of ``rungs" of a participation ladder, as outlined by Arnstein \cite{Arnstein2019}. While developers promote human-centered, participatory strategies during model design, they may not be engaging in true participation, refusing to advance through the ladder towards partnership and control \cite{Arnstein2019}. When developers consciously progress towards citizen control, stakeholders become integrated in decision-making throughout design and into implementation. Previous GROUP and CSCW research has examined how misalignment between initial model expectations and actual AI outcomes can cause tensions for workers in the public sector \cite{deng,garg,stapleton,group_25}, with HCI researchers suggesting ``early engagement of stakeholders in building equitable public technologies'' \cite{onestop}. 


As seen in our corpus, a wide range of evaluation metrics are used to assess and compare model performance (Table \ref{table:metrics}). In our coding we denote metrics such as sensitivity, recall, precision, and specificity as ``Binary Classification,'' due to their overlapping prevalence among papers (papers are multi-counted if explicitly mentioning more than one metric). The choice of evaluation metrics should arise from a combination of the technical method at hand as well as relevant sociotechnical context \cite{IEEE_standards}. Discussions between developers, stakeholders, end users, regulatory bodies, and ethics boards can further align model expectations with equitable performance when implemented in workplace settings \cite{lay_user_involve}.



In our corpus, seven papers included participatory design, a lens within HCAD. Of these seven papers, all but one (85.7\%) considered cost reduction and cost-effectiveness of AI, with the remaining one considering model cost. Whilst leveraging participatory efforts from stakeholders, these researchers have yet to consider cost from computational and model sources, instead still reflecting on cost-effectiveness of AI as a whole. Further engagement with time and labour costs, particularly of the relevant participatory pipeline, is needed to comprehensively prioritize human-centered design and a holistically cost-aware approach. Of the four deployed/implemented papers in our corpus, none consider participatory or speculative design, with only one considering sociotechnical context and a separate one acknowledging model cost. A quantitative breakdown of cost types by method and outcome can be seen in Table \ref{table:5}, while cost categories acknowledged by outcome groups can be seen in Appendix \ref{sec:tables_appendix} (Table \ref{table:cost_breakdown}).


Multiple papers in our corpus call for participatory involvement of stakeholders to voice the concerns of providers and patients, alike \cite{ACM_Gyldenkaerne,ACM_Pfohl,PubMed_Soliman}. These conversations and parallel participatory strategies provide insight into accurate model cost calculations. Suchman [2002] acknowledges the cost of integrating stakeholders and practitioners throughout design work, in both ``time and human effort" \cite{suchman}. While the system at hand benefits from additional context and stakeholder perspectives, there are time and labour costs for finding participants, bringing them to development discussions, and maintaining those connections \cite{suchman}. In clinical medicine, several metrics exist for such costs against the benefit to patients. In the United Kingdom's National Health System (NHS), metrics such as Number Needed to Treat (NNT) and the NICE Cost-Effective Threshold aim to balance the monetary or physical labour cost of a new treatment/medication against the potential negative outcomes \cite{nice}. Multiple countries have defined a Quality-Adjusted Life Year (QULY) metric \cite{qaly}, to quantify the acceptable monetary cost of an additional year of life. These frameworks have quantified the trade-off between cost and burden, allowing medicine to progress according to some cost:benefit threshold. In human-centered AI development, an existing cost framework does not exist in such prescribed values, so we instead recommend researchers approach the trade-off between participatory design cost and potential benefit to AI improvement through frameworks such as Minimal Clinically Important Difference (MCID) \cite{mcid}. Leveraging the MCID approach, quantitative values can be measured to provide a more holistic understanding of the balance between participatory effort (labour and monetary cost) and improvement in AI design (model performance through a specified evaluation metric). AI researchers and external stakeholders can be surveyed for their perspectives on the net gain from each participatory design activity, to narrow down those methods and activities that yield the largest improvement to model performance and trust, whilst decreasing known sociotechnical barriers to AI comprehensiveness and implementation.

During the model commissioning stage, speculative conversations should ask if AI models are necessary and contribute to overall positive implementations \cite{baumer_silberman}. \textbf{Aside from sustainability concerns, the tensions between model costs, stakeholder expectations, and tangible impacts of AI solutions should be questioned as realistically surmountable} \cite{baumer_silberman}. Baumer \& Silberman [2011] prompt us to reflect on alternatives for these models that rely on fewer technical resources (and thus may produce lower costs or reduce bias), 
and consider whether the problem at hand actively requires a computational solution \cite{baumer_silberman,group_kms}. Considering low-tech alternatives to AI models may increase the proportion of cost-effective technical solutions and raise awareness of model costs in the SIGCHI-health community, changing the balance we see in Figure \ref{fig:costs}. 
While this abolitionism introduced by Baumer \& Silberman [2011] is a necessary consideration in some circumstances, such as when cost- or design-related conflicts cannot be overcome, we recommend researchers consider the benefits of representative, responsible AI modeling in health \cite{chen_pubmed}, and develop AI models for health systems with the HCAD framework in mind. 

\subsection{Algorithmic Decision-Making in the Workplace} \label{sec6.NEW}
Our review of cost‑aware AI models in health systems reveals that algorithmic decision‑making is not purely technical; rather, it is embedded in organizational routines, shaped by power relations, and constrained by metrics that datafication makes visible \cite{zajac_it_2024}. GROUP and CSCW work on algorithmic management in platform labour has shown that algorithms do not simply “optimize” work; they reconfigure accountability, control, autonomy, and the ability to remain informed in current workplace activities \cite{lee_working_2015,group_kms,devansh_group}. Subsequent studies of bias in freelance marketplaces demonstrate how rating and ranking systems encode historical inequities, systematically privileging some workers over others \cite{hannak_bias_2017}. These findings resonate with our health‑system corpus: when optimization targets are framed in cost terms, the burdens of efficiency are disproportionately shifted to already‑marginalized patients and frontline staff.  

Cost‑aware health system models amplify a broader managerial turn toward datafication. Similar to the analytics deployed in higher education \cite{mcconvey2023,McConvey2024} and homelessness services \cite{Moon2024}, predictive 
tools require institutions to translate complex human needs into discrete, computable proxies. This translation foregrounds what is measurable (e.g., readmission probability, length‑of‑stay) while sidelining socio‑contextual factors that resist easy quantification. These surrounding social and environmental conditions shape a patient's situation (income security, housing quality, family support, neighborhood resources, discrimination, and cultural norms), operating outside the individual’s clinical data yet profoundly influencing health needs and outcomes. Because these conditions are hard to measure with neat numerical proxies, they are often left out of cost‑focused predictive models, leading to the potential redefining of “good care” as “least expensive care” and consequently structuring bias directly into resource‑allocation workflows \cite{guerdan_counterfactual_2023}.

Workplace studies outside the public sector further illustrate how algorithmic reasoning migrates across domains. Spektor et al.’s review of hospitality automation shows how optimization goals first framed as customer‑service improvements quickly become rationales for labour reduction and heightened surveillance \cite{spektor_charting_2023}. In policing, Haque et al. find that frontline officers with diverse AI literacy levels struggle to reconcile crime‑mapping predictions with situated expertise, prompting workflow workarounds, mistrust, and new forms of accountability tension \cite{haque_RR}. The parallel in hospitals is the growing use of electronic health records and dashboards that push staff to prioritize throughput over relational care, reproducing a gig‑economy ethos inside clinical work \cite{sun_data_2023}. These cross‑sector patterns suggest that algorithmic decision‑making is not merely applied within workplaces; it actively reshapes what counts as valuable labour and desirable outcomes in the \textbf{decision → intervention → outcome} pipeline.


\paragraph{Accounting for the full cost burden.}
Embedded in every step of that pipeline are costs accrued to different stakeholders. Administrators shoulder development, storage, and maintenance expenses; clinicians absorb training, cognitive‑load, and workflow interruption costs; and patients experience shifts in care patterns that may lengthen wait times or impact services. 
We echo recent calls for developers and procuring institutions to incorporate comprehensive cost‑of‑ownership checklists that surface both monetary and social expenditures throughout the model life cycle.

Algorithms do more than automate decisions; they institutionalize particular values by turning the metrics chosen for optimization into defacto organizational priorities. 
Technical “debiasing” in isolation cannot remedy discriminatory outcomes as bias is sociotechnical, emerging from the interplay of data quality, historical inequities, and institutional incentives. Accordingly, future research must look past accuracy benchmarks to examine how cost‑aware AI systems redistribute labor, risk, and moral responsibility within care teams. Leveraging participatory design traditions from GROUP and CSCW work can help align optimization targets with the ethical obligations owed to patients, frontline workers, and surrounding communities, ensuring that the quest for efficiency does not eclipse health care’s fundamental commitments to equity and well‑being.

\subsection{Interrogative Guidelines for Cost-Aware Model Design} \label{sec6.4}
Prompted by our literature review and previous SIGCHI research, we provide guidelines to assist future AI solutions in health systems. For those researchers developing models for health systems, or other disciplines utilizing similar human-centered algorithm techniques, we recommend the following: 


Recommendations from Our Literature Review:
\begin{itemize}
    \item \textit{Compare model expectations and costs with tangible implementation strategies.} If a model is actively required and beneficial 
    (determined through conversations with stakeholders) \cite{baumer_silberman}, then HCAD principles should be prioritized. After initial conversations with stakeholders around the expectations of a desired predictive model, specifying the desired predictors, methods, and target outcome(s), continuous prompting of \textbf{can we achieve what stakeholders need?} allows for iterative design and validation. Aligning both the technical model requirements and model costs is important for sustainable AI assistance for end users \cite{IEEE_Jiao}. Many developers endorse models labeled as ``cost-effective'' or ``reducing healthcare spending,'' while only a fraction consider model or data-driven costs (Figure \ref{fig:costs}). 
    Ensuring a common understanding of ``cost-effective'' further aligns these expectations. And as design decisions change throughout development (from new data sources, updated requirements, scalability, cost categories), all categories of model costs (Tables \ref{tab:costxHCAD},\ref{tab:framework},\ref{table:cost_breakdown}) should  be transparently re-evaluated, re-initializing the theoretical, speculative, and participatory discussions taking place \cite{Baumer17}. Careful consideration of the dynamics between cost-effectiveness, equitable model performance, and sustainability must be made, anticipating downstream impacts of each model design choice.
    \item \textit{Calculate all model costs at each stage of AI model development.} As model design decisions change throughout development (from new data sources, updated requirements, scalability), accurate model costs should be re-evaluated, re-initiating stakeholder feedback. Similar to maintaining transparency with stakeholders (and understanding that viewpoints may differ between stakeholder groups as to how ``cost-effective'' solutions may be evaluated), an increase in the amount of data, computational complexity, or model generalizability should be openly presented as more expensive to designers, and the overall budget (prompted by questions in Table \ref{tab:framework}) of the model should be re-evaluated. 
\end{itemize}

Recommendations from Our Corpus and Adjacent HCAD Research:
\begin{itemize}
    \item \textit{Emphasize co-design discussions throughout AI model development.} Integrating co-design opportunities 
    allows stakeholders to contribute to the technical expectations of model requirements, sociotechnical understandings of model needs and outputs, necessary explainability for all end users, and integration assistance \cite{ACM_Gyldenkaerne,Chui2023,PubMed_Soliman,stakeholder_discuss}. Organizing community engagement committees of patients, patient advocates, caregivers, providers, policy makers, domain experts and administrators allows model design to be tailored to the intended audience, as opposed to relying on industry or regional recommendations \cite{Chui2023,group_kms}. These meetings are already being explored in public health \cite{cifar_network}, and we encourage SIGCHI researchers to implement these conversations to discuss which predictors are most important and accessible, and further explore those risks most anticipated by users (\ref{sec6.1}) \cite{implications_elendu,JMIR_Rogers}.

    \item \textit{Understand clinician hesitations for model implementation.} Many physicians are aware that their future workflows will involve AI \cite{alzaabi_physicians_AI}, while education around AI is only partially embedded in medical school settings \cite{grunhut_AI_medicalschool}. Nurses and administrators have initially approached AI with skepticism \cite{nursing}, hesitant of those models lacking peer-reviewed validation, raising several recommended guidelines, including an easy-to-use interface, full integration into (and compatibility with) existing health tools, and benefit to patient care \cite{nursing,administrators}. AI education will need to represent a broad range of model outcomes and intentions (Figure \ref{fig:outcomes}), different implementation locations (system-level, clinic-level, physician-level) and the usage of predictor combinations (Figure \ref{fig:predictors}). 
    When discussing model development with stakeholders in community engagement committees, agenda items should include:
    \begin{itemize}
        \item Clinician hesitations that provide an acute look at optimal AI implementation into existing workflows.
        \item An accurate estimate of educational resources and time required to train clinicians.
        \item The potential impacts of - and mitigation strategies for - bias embedded throughout the AI pipeline and the resulting effects on patient care journeys and provider workflows.
    \end{itemize}
\end{itemize}

\section{Limitations and Future Work}

Restricting our search to only articles in English, that had a full technical description of predictors, methods, and outcomes, means AI models may have been missed in our review. The exclusion of survey papers and quality assessment frameworks further aligned our collected corpus within a prescribed inclusion criteria; however, may have lead to our neglect of models outside of health system purposes. We searched specific databases and selected publications with publicly available models, for a total of only 114 in our literature review. The majority of our papers were from the US (n=42), India (n=14) and China (n=13). With our framing of ``model costs,'' we examine the prevalence and trends of cost framings, excluding a consideration of exact monetary values and their comparison across currencies and year-of-development. We will further explore cost magnitude comparisons in future work, presenting a process for cost comparisons and downstream consequences of prioritizing financial design. 
All our work will be initiated, developed, and implemented with human-centered algorithm design principles \cite{Baumer17}, to provide solutions that involve sociotechnical backgrounds, appropriately meet stakeholder needs and expectations, and recognize societal, representational, and deployment bias implications. 

\section{Conclusion}

We performed a systematic literature review of 112 peer-reviewed publications and two reports outlining AI models in health systems. We explored the combinations of predictors, methods, and outcomes used, as well as cost considerations and human-centered priorities throughout development. The temporal trends of methodological choices were examined and extrapolated, and suggestions for future cost-aware modeling in health systems were outlined. We acknowledge the expanding potential of health system AI models with increased computational power and data integration and encourage designers to examine all model-related costs and downstream sustainability impacts. Providing a clear assessment of model costs and its underlying dimensions can begin to unravel the dynamics between purely `cost-effective' solutions and the subsequent readjustment of time, labour, resources, and risk across multiple health system stakeholder groups. Lastly, we present potential impacts of AI in the workplace and encourage the integration of socio-contextual factors through theoretical, speculative, and participatory approaches, towards impactful, novel, and accurate AI solutions in health systems.


\bibliographystyle{ACM-Reference-Format}
\bibliography{CHI25_HealthcareAI,kelly_references}

@article{Baumer17,
    author = {Eric PS Baumer},
    title ={Toward human-centered algorithm design},
    journal = {Big Data \& Society},
    volume = {4},
    number = {2},
    pages = {2053951717718854},
    year = {2017},
    doi = {10.1177/2053951717718854},
    URL = {https://doi.org/10.1177/2053951717718854}
}

@inproceedings{audit1,
author = {Solarova, Sara and Mosnar, Matej and Tibensky, Matus and Jakubcik, Jan and Bindas, Adrian and Liska, Simon and Hossner, Filip and Mesar\v{c}\'{\i}k, Mat\'{u}\v{s} and Srba, Ivan},
title = {The DSA's Blind Spot: Algorithmic Audit of Advertising and Minor Profiling on TikTok},
year = {2026},
isbn = {9798400725968},
publisher = {Association for Computing Machinery},
address = {New York, NY, USA},
url = {https://doi-org.myaccess.library.utoronto.ca/10.1145/3805689.3812355},
doi = {10.1145/3805689.3812355},
booktitle = {Proceedings of the 2026 ACM Conference on Fairness, Accountability, and Transparency},
pages = {4811–4835},
numpages = {25},
location = {
},
series = {FAccT '26}
}

@inproceedings{audit2,
author = {Govindu, Megha N. and Wang, Stephanie T. and Friedler, Sorelle A. and Metaxa, Dana\'{e}},
title = {Do Language Models Pass the Bechdel Test? Auditing Gender Biases in LLM-Generated Screenplays},
year = {2026},
isbn = {9798400725968},
publisher = {Association for Computing Machinery},
address = {New York, NY, USA},
url = {https://doi-org.myaccess.library.utoronto.ca/10.1145/3805689.3812208},
doi = {10.1145/3805689.3812208},
booktitle = {Proceedings of the 2026 ACM Conference on Fairness, Accountability, and Transparency},
pages = {7398–7421},
numpages = {24},
location = {
},
series = {FAccT '26}
}

@inproceedings{barrier1,
author = {Huang, Michelle and Rodr\'{\i}guez, Violeta J. and Saha, Koustuv and August, Tal},
title = {Designing Beyond Language: Sociotechnical Barriers in AI Health Technologies for Limited English Proficiency},
year = {2026},
isbn = {9798400722783},
publisher = {Association for Computing Machinery},
address = {New York, NY, USA},
url = {https://doi-org.myaccess.library.utoronto.ca/10.1145/3772318.3791091},
doi = {10.1145/3772318.3791091},
booktitle = {Proceedings of the 2026 CHI Conference on Human Factors in Computing Systems},
articleno = {771},
numpages = {18},
location = {
},
series = {CHI '26}
}

@article{barrier2,
author = {Adio, Oluwatobiloba J. and Olugbenga, Motunrayo N. and Ikwunne, Tochukwu},
title = {Bridging Gaps: Uncovering Cultural and Healthcare Barriers to Maternal Health in Rural Communities of the Global South: a Qualitative Study},
year = {2024},
issue_date = {June 2024},
publisher = {Association for Computing Machinery},
address = {New York, NY, USA},
number = {138},
issn = {1558-2337},
url = {https://doi-org.myaccess.library.utoronto.ca/10.1145/3703599.3703600},
doi = {10.1145/3703599.3703600},
journal = {SIGACCESS Access. Comput.},
month = nov,
articleno = {1},
numpages = {1}
}

@inproceedings{barrier3,
author = {Batool, Vafa},
title = {Designing Digital Health Tools to Address Stigma and Structural Barriers across the Mental Health Care Ecosystem},
year = {2026},
isbn = {9798400722813},
publisher = {Association for Computing Machinery},
address = {New York, NY, USA},
url = {https://doi-org.myaccess.library.utoronto.ca/10.1145/3772363.3799219},
doi = {10.1145/3772363.3799219},
booktitle = {Proceedings of the Extended Abstracts of the 2026 CHI Conference on Human Factors in Computing Systems},
articleno = {888},
numpages = {4},
location = {
},
series = {CHI EA '26}
}

@ARTICLE{mcid,
  title    = "Clinimetrics Corner: The Minimal Clinically Important Change
              Score ({MCID)}: A Necessary Pretense",
  author   = "Cook, Chad E",
  journal  = "J Man Manip Ther",
  volume   =  16,
  number   =  4,
  pages    = "E82--3",
  year     =  2008,
  address  = "England",
  language = "en"
}

@ARTICLE{qaly,
  title    = "Problems and solutions in calculating quality-adjusted life years
              ({QALYs})",
  author   = "Prieto, Luis and Sacrist{\'a}n, Jos{\'e} A",
  journal  = "Health Qual Life Outcomes",
  volume   =  1,
  pages    = "80",
  month    =  dec,
  year     =  2003,
  address  = "England",
  language = "en"
}

@ARTICLE{nice,
  title    = "The {NICE} cost-effectiveness threshold: what it is and what that
              means",
  author   = "McCabe, Christopher and Claxton, Karl and Culyer, Anthony J",
  journal  = "Pharmacoeconomics",
  volume   =  26,
  number   =  9,
  pages    = "733--744",
  year     =  2008,
  address  = "New Zealand",
  language = "en"
}

@inproceedings{barrier4,
author = {Dillahunt, Tawanna R and Maestre, Juan F. and Kameswaran, Vaishnav and Poon, Erica and Osorio Torres, John and Gallardo, Mia and Rasmussen, Samantha E. and Shih, Patrick C. and Bagley, Alice and Young, Samuel L. A. and Veinot, Tiffany C.},
title = {Trust, Reciprocity, and the Role of Timebanks as Intermediaries: Design Implications for Addressing Healthcare Transportation Barriers},
year = {2022},
isbn = {9781450391573},
publisher = {Association for Computing Machinery},
address = {New York, NY, USA},
url = {https://doi-org.myaccess.library.utoronto.ca/10.1145/3491102.3502494},
doi = {10.1145/3491102.3502494},
booktitle = {Proceedings of the 2022 CHI Conference on Human Factors in Computing Systems},
articleno = {521},
numpages = {22},
location = {New Orleans, LA, USA},
series = {CHI '22}
}

@inbook{audit3,
author = {Taylor, Jordan and Agnew, William and Sap, Maarten and Fox, Sarah E and Zhu, Haiyi},
title = {The Algorithmic Gaze of Image Quality Assessment: An Audit and Trace Ethnography of the LAION-Aesthetics Predictor},
year = {2026},
isbn = {9798400725968},
publisher = {Association for Computing Machinery},
address = {New York, NY, USA},
url = {https://doi.org/10.1145/3805689.3806462},
booktitle = {Proceedings of the 2026 ACM Conference on Fairness, Accountability, and Transparency},
pages = {6383–6402},
numpages = {20}
}

@inproceedings{Bharati_R_Singh_Khanna_V_C_2025, title={Low-Cost Handheld Ultrasound for Rural Healthcare: Sonolite}, ISSN={2836-1873}, url={https://ieeexplore.ieee.org/document/11089200}, DOI={10.1109/ICCSP64183.2025.11089200}, abstractNote={Access to prenatal ultrasound diagnostics remains limited in many rural areas, primarily due to the prohibitive cost of imaging systems and the lack of specialized personnel. This paper proposes Sonolite, a lightweight, AI-enabled software framework designed to facilitate fetal structure detection from ultrasound images, with a focus on deployment in low-resource environments. The system utilizes a Faster Region-Based Convolutional Neural Network (R-CNN) model with a custom ResNet-34 backbone, optimized to detect key fetal anatomical regions, including the head, abdomen, hands, and legs. The model was trained and evaluated on a curated ultrasound dataset, achieving an average accuracy over 90%. The software is designed for real-time inference and has been integrated with a mobile interface using FastAPI and Retrofit to ensure platform compatibility. A preliminary survey of rural public health centers was conducted to identify infrastructural constraints, which guided the design considerations of the framework. Future work will focus on integrating the system with handheld ultrasound devices and conducting in-field validation to assess usability and performance in practical clinical settings.}, note={ISSN: 2836-1873}, booktitle={2025 11th International Conference on Communication and Signal Processing (ICCSP)}, author={Bharati, Ritul and R, Preethi and Singh, Atharva and Khanna, Apoorv and V, Arulalan and C, Muralidharan}, year={2025}, month=june, pages={1484–1489} }

@article{Do_Allison_Nguyen_Phung_Tran_Le_Nguyen_2024, title={Applying machine learning in screening for Down Syndrome in both trimesters for diverse healthcare scenarios}, volume={10}, ISSN={2405-8440}, DOI={10.1016/j.heliyon.2024.e34476}, abstractNote={Background
This paper describes the development of low-cost, effective, non-invasive machine learning-based prediction models for Down Syndrome in the first two trimesters of pregnancy in Vietnam. These models are adaptable to different situations with limited screening capacities at community-based healthcare facilities.
Method
Ultrasound and biochemical testing alone and in combination, from both trimesters were employed to build prediction models based on k-Nearest Neighbor, Support Vector Machine, Random Forest, and Extreme Gradient Boosting algorithms.
Results
A total of 7,076 pregnant women from a single site in Northern Vietnam were included, and 1,035 had a fetus with Down Syndrome. Combined ultrasound and biochemical testing were required to achieve the highest accuracy in trimester 2, while models based only on biochemical testing performed as well as models based on combined testing during trimester 1. In trimester 1, Extreme Gradient Boosting produced the best model with 94% accuracy and 88% AUC, while Support Vector Machine produced the best model in trimester 2 with 89% accuracy and 84% AUC.
Conclusions
This study explored a range of machine learning models under different testing scenarios. Findings point to the potential feasibility of national screening, especially in settings without enough equipment and specialists, after additional model validation and fine tuning is performed.}, number={15}, journal={Heliyon}, author={Do, Huy D. and Allison, Jeroan J. and Nguyen, Hoa L. and Phung, Hai N. and Tran, Cuong D. and Le, Giang M. and Nguyen, Trang T.}, year={2024}, month=aug, pages={e34476} }

@article{Lin_Bai_Huang_Lee_Vu_Chiu_2025, title={Artificial Intelligence–Based Computerized Digit Vigilance Test in Community-Dwelling Older Adults: Development and Validation Study}, volume={13}, ISSN={2291-9694}, DOI={10.2196/73038}, abstractNote={Background: The Computerized Digit Vigilance Test (CDVT) is a well-established measure of sustained attention. However, the CDVT only measures the total reaction time and response accuracy and fails to capture other crucial attentional features such as the eye blink rate, yawns, head movements, and eye movements. Omitting such features might provide an incomplete representative picture of sustained attention.
Objective: This study aimed to develop an artificial intelligence (AI)–based Computerized Digit Vigilance Test (AI-CDVT) for older adults.
Methods: Participants were assessed by the CDVT with video recordings capturing their head and face. The Montreal Cognitive Assessment (MoCA), Stroop Color Word Test (SCW), and Color Trails Test (CTT) were also administered. The AI-CDVT was developed in three steps: (1) retrieving attentional features using OpenFace AI software (CMU MultiComp Lab), (2) establishing an AI-based scoring model with the Extreme Gradient Boosting regressor, and (3) assessing the AI-CDVT’s validity by Pearson r values and test-retest reliability by intraclass correlation coefficients (ICCs).
Results: In total, 153 participants were included. Pearson r values of the AI-CDVT with the MoCA were −0.42, −0.31 with the SCW, and 0.46–0.61 with the CTT. The ICC of the AI-CDVT was 0.78.
Conclusions: We developed an AI-CDVT, which leveraged AI to extract attentional features from video recordings and integrated them to generate a comprehensive attention score. Our findings demonstrated good validity and test-retest reliability for the AI-CDVT, suggesting its potential as a reliable and valid tool for assessing sustained attention in older adults.}, journal={JMIR Medical Informatics}, author={Lin, Gong-Hong and Bai, Dorothy and Huang, Yi-Jing and Lee, Shih-Chieh and Vu, Mai Thi Thuy and Chiu, Tsu-Hsien}, year={2025}, month=nov, pages={e73038–e73038}, language={en} }

@article{Lyth_Gialias_Husberg_Bernfort_Bjerner_Wiberg_Levin_Gustafsson_2026, title={Results from a Swedish model-based analysis of the cost-effectiveness of AI-assisted digital mammography}, volume={36}, ISSN={1432-1084}, DOI={10.1007/s00330-025-11821-9}, abstractNote={To evaluate the cost-effectiveness of AI-assisted digital mammography (AI-DM) compared to conventional biennial breast cancer digital mammography screening (cDM) with double reading of screening mammograms, and to investigate the change in cost-effectiveness based on four different sub-strategies of AI-DM.}, number={1}, journal={European Radiology}, author={Lyth, Johan and Gialias, Pantelis and Husberg, Magnus and Bernfort, Lars and Bjerner, Tomas and Wiberg, Maria Kristoffersen and Levin, Lars-Åke and Gustafsson, Håkan}, year={2026}, month=jan, pages={754–764}, language={en} }

@article{PradeepKumar_J_B_K_2025, title={Optimized anti-interference dynamic integral neural network approach for dementia prediction in health care}, volume={321}, ISSN={0950-7051}, DOI={10.1016/j.knosys.2025.113723}, abstractNote={Alzheimer’s is the most common progressive neurological illness, particularly affecting those with mild cognitive impairment. Raising awareness and promoting early diagnosis are crucial to reducing Alzheimer’s impact. Advances in machine learning offer promising tools for early prediction. In this manuscript, an Optimized Anti-Interference Dynamic Integral Neural Network Approach for Dementia Prediction in Health Care (DPHC-AIDINN) is proposed. The process begins with collecting OASIS dataset images, pre-processed using the Surface Normal Gabor Filter (SNGF) to remove noise. Shape features like Digital Bending Energy, Circularity Ratio, Rectangularity, and Convexity are then extracted. Using the Child Drawing Development Optimization Algorithm (CDDO), 12 key features are selected and fed into the Anti-Interference Dynamic Integral Neural Network (AIDINN) for dementia classification. To enhance accuracy, the Humboldt Squid Optimization Algorithm (HSOA) optimizes AIDINN, improving classification of demented and non-demented cases. This optimization boosts classification accuracy, with DPHC-AIDINN implemented in Python. The proposed strategy’s performance was evaluated using performance criteria like precision, recall, accuracy, specificity, F1-score, error rate and ROC. DPHC-AIDINN excels in early dementia prediction, achieving 99.39 % accuracy for demented and 99.85 % for non-demented cases. It enhances precision, recall, and F1-score by 97.69 %, cuts errors by 44.97 %, and improves the ROC curve by 96.59 %, outperforming existing methods, such as Classification of Alzheimer’s disease using MRI data based on Deep Learning Techniques (CAD-MRI-SVM), Classification of Vascular Dementia on magnetic resonance imaging using deep learning architectures (VD-MRI-CNN), and Classifying Dementia Severity Using MRI Radiomics Analysis of the Hippocampus and Machine Learning (CDS-MRI-XGB).}, journal={Knowledge-Based Systems}, author={Pradeep Kumar, B P and J, Ravikumar and B, Shankar B and K, Manjunath Kamath}, year={2025}, month=june, pages={113723} }

@inproceedings{Siam_Ahmed_Khan_Islam_Milon_Ahamed_Islam_2025, title={Explainable Deep Learning Models for Medical Diagnosis: Bridging the Gap between AI and Healthcare}, url={https://ieeexplore.ieee.org/document/11258368}, DOI={10.1109/ICBATS66542.2025.11258368}, abstractNote={The field of medical diagnosis using images has been transformed with the application of deep learning which has shown great accuracy in many tasks involving detection and prognosis. Nevertheless, the implementation of these models in clinical practice can be problematic due to their ‘black box’ feature since practitioners would want to understand AI recommendations and verification mechanisms. The purpose of this research is to enhance the explainability of deep learning models applied in medical diagnosis by making use of SHAP and Grad-CAM techniques. Phenomenal results were achieved with these models especially in pneumonia detection with 94% accuracy, and also provided explainable results that were sensible from a medical viewpoint. Clinicians’ studies highlighted enhanced trust and usability of the suggested models for practical medical application. Nevertheless, other challenges such as the interpretation of the models at scale, computation cost efficiency and the application domain’s requirements of interpretability continue to persist. This paper highlights the important role that explainable AI in the field of healthcare, and how it will facilitate the wide deployment of this technology to improve patient care and the concept of personalised medicine.}, booktitle={2025 3rd International Conference on Business Analytics for Technology and Security (ICBATS)}, author={Siam, Md Abubokor and Ahmed, Istiaq and Khan, Md Asif Ul Hoq and Islam, Md Ariful and Milon, Md Hasanujjaman and Ahamed, Asif and Islam, Md Zahedul}, year={2025}, month=may, pages={1–7} }

@article{Stephan_Hanselmann_Bajramovic_Schosser_Laxy_2025, title={Development and validation of prediction models for stroke and myocardial infarction in type 2 diabetes based on health insurance claims: does machine learning outperform traditional regression approaches?}, volume={24}, ISSN={1475-2840}, DOI={10.1186/s12933-025-02640-9}, abstractNote={Background
Digitalization and big health system data open new avenues for targeted prevention and treatment strategies. We aimed to develop and validate prediction models for stroke and myocardial infarction (MI) in patients with type 2 diabetes based on routinely collected high-dimensional health insurance claims and compared predictive performance of traditional regression with state-of-the-art machine learning including deep learning methods.

Methods
We used German health insurance claims from 2014 to 2019 with 287 potentially relevant literature-derived variables to predict 3-year risk of MI and stroke. Following a train-test split approach, we compared the performance of logistic methods with and without forward selection, LASSO-regularization, random forests (RF), gradient boosting (GB), multi-layer-perceptrons (MLP) and feature-tokenizer transformers (FTT). We assessed discrimination (Areas Under the Precision-Recall and Receiver-Operator Curves, AUPRC and AUROC) and calibration.

Results
Among n = 371,006 patients with type 2 diabetes (mean age: 67.2 years), 3.5% (n = 13,030) had MIs and 3.4% (n = 12,701) strokes. AUPRCs were 0.035 (MI) and 0.034 (stroke) for a null model, between 0.082 (MLP) and 0.092 (GB) for MI, and between 0.061 (MLP) and 0.073 (GB) for stoke. AUROCs were 0.5 for null models, between 0.70 (RF, MLP, FTT) and 0.71 (all other models) for MI, and between 0.66 (MLP) and 0.69 (GB) for stroke. All models were well calibrated.

Conclusions
Discrimination performance of claims-based models reached a ceiling at around 0.09 AUPRC and 0.7 AUROC. While for AUROC this performance was comparable to existing epidemiological models incorporating clinical information, comparison of other, potentially more relevant metrics, such as AUPRC, sensitivity and Positive Predictive Value was hampered by lack of reporting in the literature. The fact that machine learning including deep learning methods did not outperform more traditional approaches may suggest that feature richness and complexity were exploited before the choice of algorithm could become critical to maximize performance. Future research might focus on the impact of different feature derivation approaches on performance ceilings. In the absence of other more powerful screening alternatives, applying transparent regression-based models in routine claims, though certainly imperfect, remains a promising scalable low-cost approach for population-based cardiovascular risk prediction and stratification.

Graphical abstract




Supplementary Information
The online version contains supplementary material available at 10.1186/s12933-025-02640-9.}, journal={Cardiovascular Diabetology}, author={Stephan, Anna-Janina and Hanselmann, Michael and Bajramovic, Medina and Schosser, Simon and Laxy, Michael}, year={2025}, month=feb, pages={80} }

@article{Tung_Le_Yao_Huang_Lim_Sng_Lau_Tan_Chen_Tay_et_2025, title={Performance of Retrieval-Augmented Generation Large Language Models in Guideline-Concordant Prostate-Specific Antigen Testing: Comparative Study With Junior Clinicians}, volume={27}, ISSN={1438-8871}, DOI={10.2196/78393}, abstractNote={Background: Prostate-specific antigen (PSA) testing remains the cornerstone of early prostate cancer detection. Society guidelines for prostate cancer screening via PSA testing serve to standardize patient care and are often used by trainees, junior staff, or generalist medical practitioners to guide medical decision-making. However, adherence to guidelines is a time-consuming and challenging task, and rates of inappropriate PSA testing are high. Retrieval-augmented generation (RAG) is a method to enhance the reliability of large language models (LLMs) by grounding responses in trusted external sources.
Objective: This study aimed to evaluate a RAG-enhanced LLM system, grounded in current European Association of Urology and American Urological Association guidelines, to assess its effectiveness in providing guideline-concordant PSA screening recommendations compared to junior clinicians.
Methods: A series of 44 fictional outpatient case scenarios was developed to represent a broad spectrum of clinical presentations. A RAG pipeline was developed, comprising a life expectancy estimation module based on the Charlson Comorbidity Index, followed by LLM-generated recommendations constrained to retrieved excerpts from the European Association of Urology and American Urological Association guidelines. Five junior clinicians were tasked to provide PSA testing recommendations for the same scenarios in closed-book and open-book formats. Answers were compared for accuracy in a binomial fashion. Fleiss κ was computed to assess interrater agreement among clinicians.
Results: The RAG-LLM tool provided guideline-concordant recommendations in 95.5% (210/220) of case scenarios, compared to junior clinicians, who were correct in 62.3% (137/220) of scenarios in a closed-book format and 74.1% (163/220) of scenarios in an open-book format. The difference was statistically significant for both closed-book (P<.001) and open-book (P<.001) formats. Interrater agreement among clinicians was fair, with Fleiss κ of 0.294 and 0.321 for closed-book and open-book formats, respectively.
Conclusions: Use of RAG techniques allows LLMs to integrate complex guidelines into day-to-day medical decision-making. RAG-LLM tools in urology have the capability to enhance clinical decision-making by providing guideline-concordant recommendations for PSA testing, potentially improving the consistency of health care delivery, reducing cognitive load on clinicians, and reducing unnecessary investigations and costs. While this study used synthetic cases in a controlled simulation environment, it establishes a foundation for future validation in real-world clinical settings.}, journal={Journal of Medical Internet Research}, author={Tung, Joshua Yi Min and Le, Quan and Yao, Jinxuan and Huang, Yifei and Lim, Daniel Yan Zheng and Sng, Gerald Gui Ren and Lau, Rachel Shu En and Tan, Yu Guang and Chen, Kenneth and Tay, Kae Jack and Tan, Jen Hong and Yuen, John Shyi Peng and Cheng, Christopher Wai Sam and Ho, Henry Sun Sien}, year={2025}, month=nov, pages={e78393–e78393}, language={en} }

@inproceedings{Vusumuzi_Godwin_2025, address={New York, NY, USA}, series={icARTi ’25}, title={AI-Driven Zero-Trust Models for Blockchain-Supported Healthcare Ecosystems}, ISBN={979-8-4007-2158-8}, url={https://dl.acm.org/doi/10.1145/3774791.3774801}, DOI={10.1145/3774791.3774801}, abstractNote={The rapid adoption of mobile health (mhealth) technologies and digital healthcare platforms has amplified concerns around data privacy, security, and trust. Conventional perimeter-based security models are inadequate in such distributed ecosystems where sensitive patient information is exchanged across heterogeneous devices and networks. To address this gap, this paper proposes an AI-driven Zero Trust framework integrated with blockchain technologies for secure and privacy-preserving healthcare data sharing. The objective of this study is to explore how artificial intelligence can enhance the core principles of Zero Trust, continuous verification, least-privilege access, and adaptive risk assessment, while leveraging blockchain’s immutability and decentralization to ensure auditability and compliance. The methodology employs a Design Science Research approach, combining architectural modelling with simulation-based evaluation. The framework integrates machine learning for anomaly detection, adaptive authentication, and real-time policy enforcement, alongside a blockchain layer for immutable logging and distributed trust management. Performance and security were assessed through simulated mHealth scenarios involving wearable devices and electronic health records. The results demonstrate that the proposed framework achieves higher intrusion detection accuracy (95.2%) and superior compliance alignment (HIPAA and GDPR) compared to existing blockchain-only and AI-only healthcare models. Unlike conventional mHealth security solutions that rely on static access controls or centralized trust, the AI-driven Zero Trust framework adapts dynamically to evolving threats while maintaining verifiable audit trails through blockchain. This combination delivers stronger security resilience, reduced insider risk, and improved interoperability for real-time healthcare environments.}, booktitle={Proceedings of the 2025 International Conference on Artificial Intelligence and its Applications}, publisher={Association for Computing Machinery}, author={Vusumuzi, Malele and Godwin, Mandinyenya}, year={2025}, month=dec, pages={1–11}, collection={icARTi ’25} }

@inproceedings{Wang_Yang_Liang_Lee_Cao_2024, address={New York, NY, USA}, series={ICCBDC ’24}, title={Analyzing the Usability, Performance, and Cost-Efficiency of Deploying ML Models on BigQuery ML and Vertex AI in Google Cloud}, ISBN={979-8-4007-1725-3}, url={https://dl.acm.org/doi/10.1145/3694860.3694863}, DOI={10.1145/3694860.3694863}, abstractNote={This study compared and analyzed the usability, performance, and cost-efficiency of deploying Machine Learning (ML) models in two ML-AI platforms in Google Cloud: BigQuery ML and Vertex AI. Through the experiments with two separate cases, the analysis was conducted with MIMIC-IV datasets of hospitalized patients to deploy regression models on each platform to predict mortality and progression of diseases. The documentation, learning curve, and resource suitability of the platforms were evaluated to access their usability. The study evaluated the total running times and resource utilizations, including storage and compute, to analyze their performance and cost efficiency. The analysis results showed that BigQuery ML offers good usability with easy-to-follow documentation and a moderate learning curve for cloud users, making it more suitable for SQL-savvy users and large-scale data analytics tasks. It also showed efficient resource management and deployment despite its higher initial processing times during the training.  Vertex AI incurred higher costs due to longer training times and specific resource allocations. The findings indicate that BigQuery ML seems to be more efficient, particularly in terms of processing time and cost for the experimented clinical dataset and regression models, emphasizing its suitability for large-scale data processing tasks where efficiency is essential.}, booktitle={Proceedings of the 2024 8th International Conference on Cloud and Big Data Computing}, publisher={Association for Computing Machinery}, author={Wang, Hongyu and Yang, Jeong and Liang, Gongbo and Lee, Young and Cao, Zechun}, year={2024}, month=nov, pages={15–25}, collection={ICCBDC ’24} }

@article{Wang_Chen_Wu_Jiang_Lin_Qiu_2025, title={A robust and interpretable ensemble machine learning model for predicting healthcare insurance fraud}, volume={15}, ISSN={2045-2322}, DOI={10.1038/s41598-024-82062-x}, abstractNote={Healthcare insurance fraud imposes a significant financial burden on healthcare systems worldwide, with annual losses reaching billions of dollars. This study aims to improve fraud detection accuracy using machine learning techniques. Our approach consists of three key stages: data preprocessing, model training and integration, and result analysis with feature interpretation. Initially, we examined the dataset’s characteristics and employed embedded and permutation methods to test the performance and runtime of single models under different feature sets, selecting the minimal number of features that could still achieve high performance. We then applied ensemble techniques, including Voting, Weighted, and Stacking methods, to combine different models and compare their performances. Feature interpretation was achieved through partial dependence plots (PDP), SHAP, and LIME, allowing us to understand each feature’s impact on the predictions. Finally, we benchmarked our approach against existing studies to evaluate its advantages and limitations. The findings demonstrate improved fraud detection accuracy and offer insights into the interpretability of machine learning models in this context.}, number={1}, journal={Scientific Reports}, author={Wang, Zeyu and Chen, Xiaofang and Wu, Yiwei and Jiang, Linke and Lin, Shiming and Qiu, Gang}, year={2025}, month=jan, pages={218}, language={en} }

@inproceedings{group_25,
author = {Mathur, Niharika and Zubatiy, Tamara and Mynatt, Elizabeth D.},
title = {A Research Through Design Study on AI Explanations for Collaborative Everyday Tasks for Older Adults Aging in Place},
year = {2025},
isbn = {9798400711879},
publisher = {Association for Computing Machinery},
address = {New York, NY, USA},
url = {https://doi-org.myaccess.library.utoronto.ca/10.1145/3688828.3699640},
doi = {10.1145/3688828.3699640},
booktitle = {Companion Proceedings of the 2025 ACM International Conference on Supporting Group Work},
pages = {48–53},
numpages = {6},
location = {Hilton Head, New Jersey, USA},
series = {GROUP '25}
}

@article{lay_user_involve,
author = {Vincenzi, Beatrice and Stumpf, Simone and Taylor, Alex S. and Nakao, Yuri},
title = {Lay User Involvement in Developing Human-centric Responsible AI Systems: When and How?},
year = {2024},
issue_date = {June 2024},
publisher = {Association for Computing Machinery},
address = {New York, NY, USA},
url = {https://doi-org.myaccess.library.utoronto.ca/10.1145/3652592},
doi = {10.1145/3652592},
journal = {ACM J. Responsib. Comput.},
month = jun,
articleno = {14}
}

@misc{IEEE_standards, url={https://standards.ieee.org/wp-content/uploads/2023/07/ead-prioritizing-people-planet.pdf}, journal={Prioritizing people and planet as the metrics for responsible ai}, publisher={IEEE Standards Association}, author={IEEE, Standards Association}, title ={Prioritizing People and Planet as the Metrics for Responsible AI},year={2023}, month={Aug}}

@InProceedings{cost_ethicalAI,
author="Kemell, Kai-Kristian
and Vakkuri, Ville",
editor="Hyrynsalmi, Sami
and M{\"u}nch, J{\"u}rgen
and Smolander, Kari
and Melegati, Jorge",
title="What Is the Cost of AI Ethics? Initial Conceptual Framework and Empirical Insights",
booktitle="Software Business",
year="2024",
publisher="Springer Nature Switzerland",
address="Cham",
pages="247--262",
isbn="978-3-031-53227-6"
}

@ARTICLE{eHealth2,
  title    = "Effectiveness and cost-effectiveness of ehealth interventions in
              somatic diseases: a systematic review of systematic reviews and
              meta-analyses",
  author   = "Elbert, Niels J and van Os-Medendorp, Harmieke and van Renselaar,
              Wilco and Ekeland, Anne G and Hakkaart-van Roijen, Leona and
              Raat, Hein and Nijsten, Tamar E C and Pasmans, Suzanne G M A",
  journal  = "J Med Internet Res",
  volume   =  16,
  number   =  4,
  pages    = "e110",
  month    =  apr,
  year     =  2014,
  address  = "Canada",
  language = "en"
}

@article{Wu_Hu_Yan_Li_Li_Chen_Lin_Zeng_Li_Mo_et_2025, title={Development and Validation of a Cost-Effective Machine Learning Model for Screening Potential Rheumatoid Arthritis in Primary Healthcare Clinics}, volume={18}, ISSN={1178-7031}, DOI={10.2147/JIR.S487595}, abstractNote={OBJECTIVE: In primary healthcare, diagnosing rheumatoid arthritis (RA) is challenging due to a general lack of in-depth knowledge of RA by general practitioners (GPs) and the lack of effective tools, leading to high rates of missed diagnosis. This study focuses on a screening model for primary healthcare, aiming to improve early RA screening accuracy and efficiency at a relatively lower cost, reducing delays in GPs’ recognition of RA.
METHODS: We randomly selected 2106 participants from the RA group or combined control group (comprising healthy individuals and patients with non-RA rheumatic diseases) at Peking University Shenzhen Hospital as the developing cohort. Guided by experienced rheumatologists, we built a comprehensive database with 26 clinical features. Using 10 classical machine learning algorithms, we developed screening models. Evaluation metrics determined the best model. Employing multivariatelogistic regression results and the best-performing model to identify the least costly features, ensuring applicability in primary healthcare clinics. Subsequently, we retrained and validated our proposed model based on two primary healthcare validation cohorts.
RESULTS: In experiments, the algorithms achieved over 88% accuracy on training and test sets. Random Forest (RF) excelled with 96.20% (95% CI 95.39% to 97.02%) accuracy, 96.22% (95% CI 95.40% to 97.03%) specificity, 96.18% (95% CI 95.37% to 97.00%) sensitivity, and 96.20% (95% CI 95.39% to 97.02%) Areas Under Curves (AUC). A meticulous feature selection identified 11 key features for RA screening. In an external test on two primary healthcare datasets with these features, RF demonstrated an accuracy of 88.435% (95% CI 85.55% to 91.32%), sensitivity of 98.55% (95% CI 97.47% to 99.63%), specificity of 85.56% (95% CI 82.39% to 88.73%), and an AUC of 92.055% (95% CI 89.62% to 94.49%).
CONCLUSION: The screening model excels in automating prompt identification of RA in primary healthcare, improving the early detection of RA, and reducing delays and associated costs. Our findings contribute positively and are poised to elevate prospective RA management, fostering improvements in healthcare sector responsiveness and resource efficiency.}, journal={Journal of Inflammation Research}, author={Wu, Wenqi and Hu, Xiaohao and Yan, Linyang and Li, Zhiyin and Li, Bo and Chen, Xinpeng and Lin, Zexun and Zeng, Huiqiong and Li, Chun and Mo, Yingqian and Wu, Yalin and Wang, Qingwen}, year={2025}, pages={1511–1522}, language={eng} }

@ARTICLE{digital_eHEALTH,
  title    = "The cost-effectiveness of digital health interventions: A
              systematic review of the literature",
  author   = "Gentili, Andrea and Failla, Giovanna and Melnyk, Andriy and
              Puleo, Valeria and Tanna, Gian Luca Di and Ricciardi, Walter and
              Cascini, Fidelia",
  journal  = "Front Public Health",
  volume   =  10,
  pages    = "787135",
  month    =  aug,
  year     =  2022,
  address  = "Switzerland",
  language = "en"
}

@article{clinic_funding,
author = {Scott, John C. and Conner, Douglas A. and Venohr, Ingrid and Gade, Glenn and McKenzie, Marlene and Kramer, Andrew M. and Bryant, Lucinda and Beck, Arne},
title = {Effectiveness of a Group Outpatient Visit Model for Chronically Ill Older Health Maintenance Organization Members: A 2-Year Randomized Trial of the Cooperative Health Care Clinic},
journal = {Journal of the American Geriatrics Society},
volume = {52},
number = {9},
pages = {1463-1470},
doi = {https://doi.org/10.1111/j.1532-5415.2004.52408.x},
url = {https://agsjournals.onlinelibrary.wiley.com/doi/abs/10.1111/j.1532-5415.2004.52408.x},
year = {2004}
}

@ARTICLE{financing,
  title    = "Factors influencing cost awareness in hospitals: a scoping review",
  author   = "Fitriasari, Nikma and Prayitno, Heri and Puspandari, Diah Ayu and
              Meliala, Andreasta and Utarini, Adi",
  journal  = "BMC Health Serv Res",
  volume   =  25,
  number   =  1,
  pages    = "1570",
  month    =  dec,
  year     =  2025,
  address  = "England",
  language = "en"
}

@article{Xiao_Li_Wang_Wang_Chen_2025, title={Predictive analysis for healthcare fraud detection: Integration of probabilistic model and interpretable machine learning}, volume={719}, ISSN={0020-0255}, DOI={10.1016/j.ins.2025.122499}, abstractNote={Medical insurance fraud detection is crucial for minimizing the depletion of insurance pools. While medical expense records (MER) are valuable for this task, their limited availability is often overlooked. Extant studies directly input MER into high-dimensional machine learning (ML) models to achieve fraud detection. Another class of models generates fraud prediction based on probability modelling of samples. To explore whether these two different classes of models can cross-fertilizer each other, this paper incorporates Bayesian network (BN) into extreme gradient boosting (XGB). After obtaining the healthcare fraud predictions, this study employs these results to risk management decisions for minimizing the cost of Medical Insurance Bureaus. To make the optimal cost-based decision, we develop an instance-dependent cost-sensitive XGB (ICXGB) method. Using real-world data, we construct various variables based on the famous Recency, Frequency and Monetary (RFM) principle and empirically assess the prediction performance of probabilistic model and ML. The integrative model shows a significant improvement over BN and ICXGB. Finally, a post-hoc explanation method is adopted to quantify the contributions of the predictors and obtain some management implications.}, journal={Information Sciences}, author={Xiao, Fei and Li, Han-xiong and Wang, Xiao-kang and Wang, Jian-qiang and Chen, Shui-xia}, year={2025}, month=nov, pages={122499} }

@article{Zhang_Zhu_Chen_Wang_2025, title={Predicting high-need high-cost pediatric hospitalized patients in China based on machine learning methods}, volume={15}, ISSN={2045-2322}, DOI={10.1038/s41598-025-99546-z}, abstractNote={Rapidly increasing healthcare spending globally is significantly driven by high-need, high-cost (HNHC) patients, who account for the top 5% of annual healthcare costs but over half of total expenditures. The programs targeting existing HNHC patients have shown limited long-term impact, and research predicting HNHC pediatric patients in China is limited. There is an urgent need to establish a specific, valid, and reliable prediction model using machine-learning-based methods to identify potential HNHC pediatric patients and implement proactive interventions before high costs arise. This study used a 7-year retrospective cohort dataset from two administrative databases in Shanghai, covering pediatric patients under 18 years. The machine-learning-based models were developed to predict HNHC status using logistic regression, k-nearest neighbors (KNN), random forest (RF), multi-layer perceptron (MLP), and Naive Bayes. This study divided the data from 2021–2022 into 70:30 as a training set and a test set, with the internal class balancing approach of the Synthetic Minority Over-sampling Technique (SMOTE). A grid search strategy was employed with k-fold cross-validation to optimize hyperparameters. Model performance was assessed by 5 metrics: Receiver Operating Characteristic-Area Under Curve (ROC-AUC), accuracy, sensitivity, specificity, and F1 score. The external validation from 2022–2023 data and the internal validation using different train-test ratios (80:20 and 90:10) were used to assess the robustness of the trained models. Among the 91,882 hospitalized children included in 2021, significant differences were found in socioeconomics, disease, healthcare service utilization, previous healthcare expenditure, and hospital characteristics between the HNHC and non-HNHC groups. The hospitalization costs for HNHC pediatric patients accounted for over 35% of total spending. The MLP model demonstrated the highest predictive performance (ROC-AUC: 0.872), followed by RF (0.869), KNN (0.836), and naive Bayes (0.828). The most important predictive factors included length of stay, number of hospitalizations, previous HNHC status, age, and presence of Top 20 HNHC diseases. MLP showed robustness as the most efficient model in external validation (ROC-AUC: 0.843) and internal validation using different train-test ratios (ROC-AUC: 0.826 in 80:20 ratio; 0.807 in 90:10 ratio). Machine learning models, particularly MLP, effectively predict HNHC pediatric patients, providing a basis for early identification of HNHC and proactive healthcare interventions into clinical practice. This approach can also assist policymakers and payers in optimizing healthcare resource allocation, controlling healthcare costs, and improving patient outcomes.}, number={1}, journal={Scientific Reports}, author={Zhang, Peng and Zhu, Bifan and Chen, Xing and Wang, Linan}, year={2025}, month=may, pages={16006}, language={en} }

@inproceedings{pacs,
author = {Lundberg, Nina},
title = {Impacts of PACS on radiological work},
year = {1999},
isbn = {1581130651},
publisher = {Association for Computing Machinery},
address = {New York, NY, USA},
url = {https://doi-org.myaccess.library.utoronto.ca/10.1145/320297.320316},
doi = {10.1145/320297.320316},
booktitle = {Proceedings of the 1999 ACM International Conference on Supporting Group Work},
pages = {169–178},
numpages = {10},
location = {Phoenix, Arizona, USA},
series = {GROUP '99}
}

@inproceedings{group_kms,
author = {Hoffmann, Marcel and Loser, Kai-Uwe and Walter, Thomas and Herrmann, Thomas},
title = {A design process for embedding knowledge management in everyday work},
year = {1999},
isbn = {1581130651},
publisher = {Association for Computing Machinery},
address = {New York, NY, USA},
url = {https://doi-org.myaccess.library.utoronto.ca/10.1145/320297.320332},
doi = {10.1145/320297.320332},
booktitle = {Proceedings of the 1999 ACM International Conference on Supporting Group Work},
pages = {296–305},
numpages = {10},
location = {Phoenix, Arizona, USA},
series = {GROUP '99}
}

@inproceedings{patient_physician,
author = {Bardram, Jakob E. and Bossen, Claus and Thomsen, Anders},
title = {Designing for transformations in collaboration: a study of the deployment of homecare technology},
year = {2005},
isbn = {1595932232},
publisher = {Association for Computing Machinery},
address = {New York, NY, USA},
url = {https://doi-org.myaccess.library.utoronto.ca/10.1145/1099203.1099254},
doi = {10.1145/1099203.1099254},
booktitle = {Proceedings of the 2005 ACM International Conference on Supporting Group Work},
pages = {294–303},
numpages = {10},
location = {Sanibel Island, Florida, USA},
series = {GROUP '05}
}

@inproceedings{COMPASS,
author = {Chui, Victoria and McConvey, Kelly and Moon, Erina Seh-Young and Ghai, Maya and Guha, Shion},
title = {Towards Sustainable Community-Designed AI Systems in the Public Sector},
year = {2025},
isbn = {9798400714849},
publisher = {Association for Computing Machinery},
address = {New York, NY, USA},
url = {https://doi-org.myaccess.library.utoronto.ca/10.1145/3715335.3737683},
doi = {10.1145/3715335.3737683},
booktitle = {Proceedings of the 2025 ACM SIGCAS/SIGCHI Conference on Computing and Sustainable Societies},
pages = {837–840},
numpages = {4},
location = {
},
series = {COMPASS '25}
}

@article{ACM_AlRamahi,
author = {Al-Ramahi, Mohammad and Noteboom, Cherie},
title = {Mining User-generated Content of Mobile Patient Portal: Dimensions of User Experience},
year = {2020},
issue_date = {September 2020},
publisher = {Association for Computing Machinery},
address = {New York, NY, USA},
volume = {3},
number = {3},
url = {https://doi.org/10.1145/3394831},
doi = {10.1145/3394831},
journal = {Trans. Soc. Comput.},
month = {jun},
articleno = {15},
numpages = {24}
}

@inproceedings{ACM_Banerjee,
author = {Banerjee, Shibabroto and Sood, Pourush and Ghose, Sujoy and Das, Partha Pratim},
title = {Coronary Artery Disease Classification from Photoplethysmographic Signals},
year = {2020},
isbn = {9781450377768},
publisher = {Association for Computing Machinery},
address = {New York, NY, USA},
url = {https://doi.org/10.1145/3418094.3418116},
doi = {10.1145/3418094.3418116},
booktitle = {Proceedings of the 4th International Conference on Medical and Health Informatics},
pages = {246–251},
numpages = {6},
location = {Kamakura City, Japan},
series = {ICMHI '20}
}

@inproceedings{ACM_Barillaro,
author = {Barillaro, Luca and Agapito, Giuseppe and Cannataro, Mario},
title = {Scalable deep learning for healthcare: methods and applications},
year = {2022},
isbn = {9781450393867},
publisher = {Association for Computing Machinery},
address = {New York, NY, USA},
url = {https://doi.org/10.1145/3535508.3545590},
doi = {10.1145/3535508.3545590},
booktitle = {Proceedings of the 13th ACM International Conference on Bioinformatics, Computational Biology and Health Informatics},
articleno = {73},
numpages = {8},
location = {Northbrook, Illinois},
series = {BCB '22}
}

@article{ACM_Bartenschlager,
author = {Bartenschlager, Christina C. and Ebel, Stefanie S. and Kling, Sebastian and Vehreschild, Janne and Zabel, Lutz T. and Spinner, Christoph D. and Schuler, Andreas and Heller, Axel R. and Borgmann, Stefan and Hoffmann, Reinhard and Rieg, Siegbert and Messmann, Helmut and Hower, Martin and Brunner, Jens O. and Hanses, Frank and R\"{o}mmele, Christoph},
title = {COVIDAL: A Machine Learning Classifier for Digital COVID-19 Diagnosis in German Hospitals},
year = {2023},
issue_date = {June 2023},
publisher = {Association for Computing Machinery},
address = {New York, NY, USA},
volume = {14},
number = {2},
issn = {2158-656X},
url = {https://doi.org/10.1145/3567431},
doi = {10.1145/3567431},
journal = {ACM Trans. Manage. Inf. Syst.},
month = {mar},
articleno = {14},
numpages = {16}
}

@inproceedings{ACM_Gyldenkaerne,
author = {H. Gyldenkaerne, Christopher and From, Gustav and M\o{}nsted, Troels and Simonsen, Jesper},
title = {PD and The Challenge of AI in Health-Care},
year = {2020},
isbn = {9781450376068},
publisher = {Association for Computing Machinery},
address = {New York, NY, USA},
url = {https://doi.org/10.1145/3384772.3385138},
doi = {10.1145/3384772.3385138},
booktitle = {Proceedings of the 16th Participatory Design Conference 2020 - Participation(s) Otherwise - Volume 2},
pages = {26–29},
numpages = {4},
location = {Manizales, Colombia},
series = {PDC '20}
}

@inproceedings{ACM_Hon,
author = {Hon, Chun Pan and Pereira, Mayana and Sushmita, Shanu and Teredesai, Ankur and De Cock, Martine},
title = {Risk Stratification for Hospital Readmission of Heart Failure Patients: A Machine Learning Approach},
year = {2016},
isbn = {9781450342254},
publisher = {Association for Computing Machinery},
address = {New York, NY, USA},
url = {https://doi.org/10.1145/2975167.2985648},
doi = {10.1145/2975167.2985648},
booktitle = {Proceedings of the 7th ACM International Conference on Bioinformatics, Computational Biology, and Health Informatics},
pages = {491–492},
numpages = {2},
location = {Seattle, WA, USA},
series = {BCB '16}
}

@inproceedings{ACM_HuangSong,
author = {Huang, Yi and Song, Insu},
title = {Indexing Biosignal for Integrated Health Social Networks},
year = {2020},
isbn = {9781450372992},
publisher = {Association for Computing Machinery},
address = {New York, NY, USA},
url = {https://doi.org/10.1145/3375923.3375936},
doi = {10.1145/3375923.3375936},
booktitle = {Proceedings of the 2019 6th International Conference on Biomedical and Bioinformatics Engineering},
pages = {133–141},
numpages = {9},
location = {Shanghai, China},
series = {ICBBE '19}
}

@ARTICLE{IEEE_Jiao,
  author={Jiao, Weiqi and Zhang, Xuan and D’Souza, Fabian},
  journal={IEEE Access}, 
  title={The Economic Value and Clinical Impact of Artificial Intelligence in Healthcare: A Scoping Literature Review}, 
  year={2023},
  volume={11},
  number={},
  pages={123445-123457},
  doi={10.1109/ACCESS.2023.3327905}}

@inproceedings{ACM_Jo,
author = {Jo, Eunkyung and Jeong, Yuin and Park, Sohyun and Epstein, Daniel A. and Kim, Young-Ho},
title = {Understanding the Impact of Long-Term Memory on Self-Disclosure with Large Language Model-Driven Chatbots for Public Health Intervention},
year = {2024},
isbn = {9798400703300},
publisher = {Association for Computing Machinery},
address = {New York, NY, USA},
url = {https://doi.org/10.1145/3613904.3642420},
doi = {10.1145/3613904.3642420},
booktitle = {Proceedings of the CHI Conference on Human Factors in Computing Systems},
articleno = {440},
numpages = {21},
location = {Honolulu, HI, USA},
series = {CHI '24}
}

@article{ACM_Kudyba,
author = {Kudyba, Stephan and Hamar, G. Brent and Gandy, William M.},
title = {Enhancing efficiency in the health care industry},
year = {2005},
issue_date = {December 2005},
publisher = {Association for Computing Machinery},
address = {New York, NY, USA},
volume = {48},
number = {12},
issn = {0001-0782},
url = {https://doi.org/10.1145/1101779.1101785},
doi = {10.1145/1101779.1101785},
journal = {Commun. ACM},
month = {dec},
pages = {107–110},
numpages = {4}
}

@article{ACM_Lv,
author = {Lv, Zhihan and Yu, Zengchen and Xie, Shuxuan and Alamri, Atif},
title = {Deep Learning-based Smart Predictive Evaluation for Interactive Multimedia-enabled Smart Healthcare},
year = {2022},
issue_date = {February 2022},
publisher = {Association for Computing Machinery},
address = {New York, NY, USA},
volume = {18},
number = {1s},
issn = {1551-6857},
url = {https://doi.org/10.1145/3468506},
doi = {10.1145/3468506},
journal = {ACM Trans. Multimedia Comput. Commun. Appl.},
month = {jan},
articleno = {43},
numpages = {20}
}

@inproceedings{ACM_Mahmud,
author = {Mahmud, S. M. Hasan and Hossin, Md Altab and Ahmed, Md. Razu and Noori, Sheak Rashed Haider and Sarkar, Md Nazirul Islam},
title = {Machine Learning Based Unified Framework for Diabetes Prediction},
year = {2018},
isbn = {9781450365826},
publisher = {Association for Computing Machinery},
address = {New York, NY, USA},
url = {https://doi.org/10.1145/3297730.3297737},
doi = {10.1145/3297730.3297737},
booktitle = {Proceedings of the 2018 International Conference on Big Data Engineering and Technology},
pages = {46–50},
numpages = {5},
location = {Chengdu, China},
series = {BDET '18}
}

@inproceedings{ACM_Oota,
author = {Oota, Subba Reddy and Rahman, Nafisur and Mohammed, Shahid Saleem and Galitz, Jeffrey and Liu, Minghsun},
title = {Wound and Episode Level Readmission Risk or Weeks to Readmit: Why do patients get readmitted? How long does it take for a patient to get readmitted?},
year = {2021},
isbn = {9781450388177},
publisher = {Association for Computing Machinery},
address = {New York, NY, USA},
url = {https://doi.org/10.1145/3430984.3431005},
doi = {10.1145/3430984.3431005},
booktitle = {Proceedings of the 3rd ACM India Joint International Conference on Data Science \& Management of Data (8th ACM IKDD CODS \& 26th COMAD)},
pages = {359–365},
numpages = {7},
location = {Bangalore, India},
series = {CODS-COMAD '21}
}

@inproceedings{ACM_Pfohl,
author = {Pfohl, Stephen and Xu, Yizhe and Foryciarz, Agata and Ignatiadis, Nikolaos and Genkins, Julian and Shah, Nigam},
title = {Net benefit, calibration, threshold selection, and training objectives for algorithmic fairness in healthcare},
year = {2022},
isbn = {9781450393522},
publisher = {Association for Computing Machinery},
address = {New York, NY, USA},
url = {https://doi.org/10.1145/3531146.3533166},
doi = {10.1145/3531146.3533166},
booktitle = {Proceedings of the 2022 ACM Conference on Fairness, Accountability, and Transparency},
pages = {1039–1052},
numpages = {14},
location = {Seoul, Republic of Korea},
series = {FAccT '22}
}

@INPROCEEDINGS{IEEE_Singh,
  author={Singh, Mahender and Mittal, Manisha and Dewan, Prachi and Kaur, Avneet and Kaur, Gurleen and Gupta, Ankur},
  booktitle={2024 11th International Conference on Computing for Sustainable Global Development (INDIACom)}, 
  title={From Text to Treatment: An Overview of Artificial Intelligence Chatbots in Healthcare}, 
  year={2024},
  volume={},
  number={},
  pages={690-696},
  doi={10.23919/INDIACom61295.2024.10498497}}

@inproceedings{ACM_Singh,
author = {Singh, Himanshu and Moirangthem, Biken and Pratap, Ajay and Kumari, Shilpi and Kumar, Abhishek and K. Das, Sajal},
title = {Splitfed-based Patient Severity Prediction and Utility Maximization in Industrial Healthcare 4.0},
year = {2024},
isbn = {9798400716737},
publisher = {Association for Computing Machinery},
address = {New York, NY, USA},
url = {https://doi.org/10.1145/3631461.3631953},
doi = {10.1145/3631461.3631953},
booktitle = {Proceedings of the 25th International Conference on Distributed Computing and Networking},
pages = {388–393},
numpages = {6},
location = {Chennai, India},
series = {ICDCN '24}
}

@ARTICLE{digital_labour_3,
  title    = "Digital labour and development: impacts of global digital labour
              platforms and the gig economy on worker livelihoods",
  author   = "Graham, Mark and Hjorth, Isis and Lehdonvirta, Vili",
  journal  = "Transfer (Bruss)",
  volume   =  23,
  number   =  2,
  pages    = "135--162",
  month    =  mar,
  year     =  2017,
  address  = "England",
  language = "en"
}

@ARTICLE{safehome,
  title    = "{SAFE@HOME}: Cost analysis of a new care pathway including a
              digital health platform for women at increased risk of
              preeclampsia",
  author   = "van den Heuvel, Josephus F M and van Lieshout, Christiaan and
              Franx, Arie and Frederix, Geert and Bekker, Mireille N",
  journal  = "Pregnancy Hypertension",
  volume   =  24,
  pages    = "118--123",
  month    =  jun,
  year     =  2021
}

@inproceedings{ACM_Zhang_Explain,
author = {Zhang, Alwin Yaoxian and Lam, Sean Shao Wei and Ong, Marcus Eng Hock and Tang, Phua Hwee and Chan, Ling Ling},
title = {Explainable AI: Classification of MRI Brain Scans Orders for Quality Improvement},
year = {2019},
isbn = {9781450370165},
publisher = {Association for Computing Machinery},
address = {New York, NY, USA},
url = {https://doi.org/10.1145/3365109.3368791},
doi = {10.1145/3365109.3368791},
booktitle = {Proceedings of the 6th IEEE/ACM International Conference on Big Data Computing, Applications and Technologies},
pages = {95–102},
numpages = {8},
location = {Auckland, New Zealand},
series = {BDCAT '19}
}

@INPROCEEDINGS{IEEE_Ezenkwu,
  author={Ezenkwu, Chinedu Pascal and Stephen, Bliss Utibe-Abasi and Affiah, Iniabasi and Daniel, Betabasi},
  booktitle={2023 IEEE AFRICON}, 
  title={A Green AI Model Selection Strategy for Computer-Aided Mpox Detection}, 
  year={2023},
  volume={},
  number={},
  pages={1-6},
  doi={10.1109/AFRICON55910.2023.10293707}}

@ARTICLE{IEEE_Gupta,
  author={Gupta, Rajesh and Shukla, Arpit and Tanwar, Sudeep},
  journal={IEEE Transactions on Network Science and Engineering}, 
  title={BATS: A Blockchain and AI-Empowered Drone-Assisted Telesurgery System Towards 6G}, 
  year={2021},
  volume={8},
  number={4},
  pages={2958-2967},
  doi={10.1109/TNSE.2020.3043262}}

@ARTICLE{IEEE_Lee,
  author={Lee, Hyo Kyung and Jin, Rebecca and Feng, Yuan and Bain, Philip A. and Goffinet, Jo and Baker, Christine and Li, Jingshan},
  journal={IEEE Journal of Biomedical and Health Informatics}, 
  title={An Analytical Framework for TJR Readmission Prediction and Cost-Effective Intervention}, 
  year={2019},
  volume={23},
  number={4},
  pages={1760-1772},
  doi={10.1109/JBHI.2018.2859581}}

@INPROCEEDINGS{IEEE_MS_Thisin,
  author={M.S, Saravanan and Thisin, Syed},
  booktitle={2024 International Conference on Advances in Data Engineering and Intelligent Computing Systems (ADICS)}, 
  title={Integrating AI and IoT for Enhanced Predictive Healthcare Monitoring: A Comprehensive Study on Breast Cancer Patient-Centric Approach}, 
  year={2024},
  volume={},
  number={},
  pages={1-5},
  doi={10.1109/ADICS58448.2024.10533651}}

@ARTICLE{IEEE_Massa,
  author={Massa, Silvia Maria and Riboni, Daniele and Nazarpour, Kianoush},
  journal={IEEE Transactions on Consumer Electronics}, 
  title={Explainable AI-Powered Graph Neural Networks for HD EMG-Based Gesture Intention Recognition}, 
  year={2024},
  volume={70},
  number={1},
  pages={4499-4506},
  doi={10.1109/TCE.2023.3333421}}

@INPROCEEDINGS{IEEE_Mistry,
  author={Mistry, Chinmay and Thakker, Urvish and Gupta, Rajesh and Obaidat, Mohammad S. and Tanwar, Sudeep and Kumar, Neeraj and Rodrigues, Joel J. P. C.},
  booktitle={ICC 2021 - IEEE International Conference on Communications}, 
  title={MedBlock: An AI-enabled and Blockchain-driven Medical Healthcare System for COVID-19}, 
  year={2021},
  volume={},
  number={},
  pages={1-6},
  doi={10.1109/ICC42927.2021.9500397}}

@INPROCEEDINGS{IEEE_Pius,
  author={Pius, Assumpta Mbatha and Ogada, Kennedy and Mwalili, Tobias},
  booktitle={2021 22nd International Arab Conference on Information Technology (ACIT)}, 
  title={Supervised Machine Learning Modelling of Demand for Outpatient Health-Care Services in Kenya using Artificial Neural Networks and Regression Decision Trees}, 
  year={2021},
  volume={},
  number={},
  pages={1-7},
  doi={10.1109/ACIT53391.2021.9677245}}

@INPROCEEDINGS{IEEE_Sinha,
  author={Sinha, Priyanshu and Gichoya, Judy W. and Purkayastha, Saptarshi},
  booktitle={2022 IEEE Healthcare Innovations and Point of Care Technologies (HI-POCT)}, 
  title={Leapfrogging Medical AI in Low-Resource Contexts Using Edge Tensor Processing Unit}, 
  year={2022},
  volume={},
  number={},
  pages={67-70},
  doi={10.1109/HI-POCT54491.2022.9744071}}

@Article{JMIR_Mens,
author="Van Mens, Kasper
and Lokkerbol, Joran
and Wijnen, Ben
and Janssen, Richard
and de Lange, Robert
and Tiemens, Bea",
title="Predicting Undesired Treatment Outcomes With Machine Learning in Mental Health Care: Multisite Study",
journal="JMIR Med Inform",
year="2023",
month="Aug",
day="23",
volume="11",
pages="e44322",
issn="2291-9694",
doi="10.2196/44322",
url="https://medinform.jmir.org/2023/1/e44322",
url="https://doi.org/10.2196/44322",
url="http://www.ncbi.nlm.nih.gov/pubmed/37623374"
}

@Article{JMIR_Oates,
author="Oates, John
and Shafiabady, Niusha
and Ambagtsheer, Rachel
and Beilby, Justin
and Seiboth, Chris
and Dent, Elsa",
title="Evolving Hybrid Partial Genetic Algorithm Classification Model for Cost-effective Frailty Screening: Investigative Study",
journal="JMIR Aging",
year="2022",
month="Oct",
day="7",
volume="5",
number="4",
pages="e38464",
issn="2561-7605",
doi="10.2196/38464",
url="https://aging.jmir.org/2022/4/e38464",
url="https://doi.org/10.2196/38464",
url="http://www.ncbi.nlm.nih.gov/pubmed/36206042"
}

@Article{JMIR_Rajkumar,
author="Rajkumar, Ethan
and Nguyen, Kevin
and Radic, Sandra
and Paa, Jubelle
and Geng, Qiyang",
title="Machine Learning and Causal Approaches to Predict Readmissions and Its Economic Consequences Among Canadian Patients With Heart Disease: Retrospective Study",
journal="JMIR Form Res",
year="2023",
month="May",
day="26",
volume="7",
pages="e41725",
issn="2561-326X",
doi="10.2196/41725",
url="https://formative.jmir.org/2023/1/e41725",
url="https://doi.org/10.2196/41725",
url="http://www.ncbi.nlm.nih.gov/pubmed/37234042"
}

@Article{JMIR_Rogers,
author="Rogers, Parker
and Boussina, Aaron E
and Shashikumar, Supreeth P
and Wardi, Gabriel
and Longhurst, Christopher A
and Nemati, Shamim",
title="Optimizing the Implementation of Clinical Predictive Models to Minimize National Costs: Sepsis Case Study",
journal="J Med Internet Res",
year="2023",
month="Feb",
day="13",
volume="25",
pages="e43486",
issn="1438-8871",
doi="10.2196/43486",
url="https://www.jmir.org/2023/1/e43486",
url="https://doi.org/10.2196/43486",
url="http://www.ncbi.nlm.nih.gov/pubmed/36780203"
}

@inproceedings{huber_public,
author = {Huber, Linda and Singh, Anubha and Dombrowski, Lynn and Guha, Shion and Hardy, Jean and Holten M\o{}ller, Naja},
title = {Datafication Dilemmas: Data Governance in the Public Interest},
year = {2024},
isbn = {9798400711145},
publisher = {Association for Computing Machinery},
address = {New York, NY, USA},
url = {https://doi.org/10.1145/3678884.3689137},
doi = {10.1145/3678884.3689137},
booktitle = {Companion Publication of the 2024 Conference on Computer-Supported Cooperative Work and Social Computing},
pages = {114–116},
numpages = {3},
location = {San Jose, Costa Rica},
series = {CSCW Companion '24}
}

@online{McKinsey,
  author = {Al-Haque, Shahed and Khanna, Vipul and Mandal, Suman and Rayasam, Mahi and Singh, Pooja},
  title = {AI ushers in next-gen prior authorization in healthcare},
  year = 2022,
  url = {https://www.mckinsey.com/industries/healthcare/our-insights/ai-ushers-in-next-gen-prior-authorization-in-healthcare#/},
  urldate = {2022-04-19}
}

@ARTICLE{Nature_Liu,
  title    = "Early prediction of diagnostic-related groups and estimation of hospital cost by processing clinical notes",
  author   = "Liu, Jinghui and Capurro, Daniel and Nguyen, Anthony and Verspoor, Karin",
  journal  = "npj Digital Medicine",
  volume   =  4,
  number   =  1,
  pages    = "103",
  month    =  jul,
  year     =  2021
}

@ARTICLE{Nature_Zhao,
  title    = "Assessment of medication self-administration using artificial
              intelligence",
  author   = "Zhao, Mingmin and Hoti, Kreshnik and Wang, Hao and Raghu,
              Aniruddh and Katabi, Dina",
  journal  = "Nature Medicine",
  volume   =  27,
  number   =  4,
  pages    = "727--735",
  month    =  apr,
  year     =  2021
}

@Article{PubMed_Feretzakis,
AUTHOR = {Feretzakis, Georgios and Loupelis, Evangelos and Sakagianni, Aikaterini and Kalles, Dimitris and Martsoukou, Maria and Lada, Malvina and Skarmoutsou, Nikoletta and Christopoulos, Constantinos and Valakis, Konstantinos and Velentza, Aikaterini and Petropoulou, Stavroula and Michelidou, Sophia and Alexiou, Konstantinos},
TITLE = {Using Machine Learning Techniques to Aid Empirical Antibiotic Therapy Decisions in the Intensive Care Unit of a General Hospital in Greece},
JOURNAL = {Antibiotics},
VOLUME = {9},
YEAR = {2020},
NUMBER = {2},
ARTICLE-NUMBER = {50},
URL = {https://www.mdpi.com/2079-6382/9/2/50},
PubMedID = {32023854},
ISSN = {2079-6382},
DOI = {10.3390/antibiotics9020050}
}

@Article{model_cost_application_4,
author={Lee, Shin-Jye
and Tseng, Ching-Hsun
and Lin, G. T. -- R.
and Yang, Yun
and Yang, Po
and Muhammad, Khan
and Pandey, Hari Mohan},
title={A dimension-reduction based multilayer perception method for supporting the medical decision making},
journal={Pattern Recognition Letters},
year={2020},
month={Mar},
day={01},
volume={131},
pages={15-22},
issn={0167-8655},
url={https://www.sciencedirect.com/science/article/pii/S0167865519303484}
}

@Misc{model_cost_application_3,
author={Garc{\'i}a M{\'a}rquez, Fausto P.
and Segovia Ram{\'i}rez, Isaac
and Pliego Marug{\'a}n, Alberto},
title={Decision Making using Logical Decision Tree and Binary Decision Diagrams: A Real Case Study of Wind Turbine Manufacturing},
year={2019},
volume={12},
number={9},
issn={1996-1073},
JOURNAL = {Energies},
doi={10.3390/en12091753},
url={https://doi.org/10.3390/en12091753}
}

@article{PubMed_Schwendicke,
author = {F. Schwendicke and J. Cejudo Grano de Oro and A. Garcia Cantu and H. Meyer-Lueckel and A. Chaurasia and J. Krois},
title ={Artificial Intelligence for Caries Detection: Value of Data and Information},

journal = {Journal of Dental Research},
volume = {101},
number = {11},
pages = {1350-1356},
year = {2022},
doi = {10.1177/00220345221113756},
    note ={PMID: 35996332},

URL = { 
    
        https://doi.org/10.1177/00220345221113756
    
    

}
}

@Article{model_cost_application_2,
author={Petitet, Pierre
and Attaallah, Bahaaeddin
and Manohar, Sanjay G.
and Husain, Masud},
title={The computational cost of active information sampling before decision-making under uncertainty},
journal={Nature Human Behaviour},
year={2021},
month={Jul},
day={01},
volume={5},
number={7},
pages={935-946},
issn={2397-3374},
doi={10.1038/s41562-021-01116-6},
url={https://doi.org/10.1038/s41562-021-01116-6}
}

@article{model_cost_application_1,
author = {Bruns, Morgan and Paredis, Christiaan and Ferson, Scott},
year = {2012},
month = {08},
pages = {},
title = {Computational Methods for Decision Making Based on Imprecise Information}
}

@ARTICLE{rua_disability_RR,
  author={Williams, Rua M. and Smarr, Simone and Prioleau, Diandra and Gilbert, Juan E.},
  journal={IEEE Transactions on Technology and Society}, 
  title={Oh No, Not Another Trolley! On the Need for a Co-Liberative Consciousness in CS Pedagogy}, 
  year={2022},
  volume={3},
  number={1},
  pages={67-74},
  doi={10.1109/TTS.2021.3084913}}

@article{Zahlan_2023, title={Artificial intelligence innovation in healthcare: Literature review, exploratory analysis, and future research}, volume={74}, ISSN={0160-791X}, DOI={10.1016/j.techsoc.2023.102321}, abstractNote={Artificial intelligence (AI) innovation in healthcare has emerged as an increasingly significant area of research. AI, digital data collection, and computer infrastructure advancements have empowered humans to address complex healthcare challenges. This study conducts a systematic literature review (SLR) of peer-reviewed journal articles at the intersection of AI, innovation, and healthcare to offer research directions for scholars and leaders in healthcare management. To achieve this, the systematic review identified and analyzed 378 published studies on AI innovation in healthcare. Evaluating these publications based on inclusion and exclusion criteria yielded 75 studies ultimately selected for comprehensive analysis. This research adds to the scope of previous investigations by aiming to 1) emphasize the most crucial AI-based healthcare applications, 2) explore challenges associated with AI integration in healthcare, and 3) examine student adoption and incorporation of AI into existing healthcare curricula. We also conducted an exploratory study of over 2700 AI-enabled healthcare startups worldwide to supplement our literature review. The SLR reveals several gaps within the research scope and proposes corresponding future research directions. These future research directions will assist researchers and enable healthcare professionals to develop legislation that accelerates the adoption of AI solutions in healthcare, ultimately enhancing public access to efficient and effective healthcare services.}, journal={Technology in Society}, author={Zahlan, Ahmed and Ranjan, Ravi Prakash and Hayes, David}, year={2023}, month=aug, pages={102321} }

@article{Vishwakarma, title={Application of artificial intelligence for resilient and sustainable healthcare system: systematic literature review and future research directions}, volume={0}, ISSN={0020-7543}, DOI={10.1080/00207543.2023.2188101}, abstractNote={Recent years have witnessed increased pressure across the global healthcare system during the COVID-19 pandemic. The COVID-19 pandemic shattered existing healthcare operations and taught us the importance of a resilient and sustainable healthcare system. Digitisation, specifically adoption of Artificial Intelligence (AI) has positively contributed to developing a resilient healthcare system in recent past. To understand how AI contributes to building a resilient and sustainable healthcare system, this study based on systematic literature review of 89 articles extracted from Scopus and Web of Science databases is conducted. The study is organised around several key themes such as applications, benefits, and challenges of using AI technology in healthcare sector. It is observed that AI has wide applications in radiology, surgery, medical, research, and development of healthcare sector. Based on the analysis, a research framework is proposed using an extended Antecedents, Practices, and Outcomes (APO) framework. This framework comprises AI applications’ antecedents, practices, and outcomes for building a resilient and sustainable healthcare system. Consequently, three propositions are drawn in this study. Furthermore, our study has adopted the theory, context and methodology (TCM) framework to provide future research directions, which can be used as a reference point for future studies.}, number={0}, journal={International Journal of Production Research}, publisher={Taylor & Francis}, author={Vishwakarma, Laxmi Pandit and Singh, Rajesh Kr and Mishra, Ruchi and Kumari, Archana}, pages={1–23}, year={2023}}

@article{ali_2023, title={The enlightening role of explainable artificial intelligence in medical \& healthcare domains: A systematic literature review}, volume={166}, ISSN={0010-4825}, DOI={10.1016/j.compbiomed.2023.107555}, abstractNote={In domains such as medical and healthcare, the interpretability and explainability of machine learning and artificial intelligence systems are crucial for building trust in their results. Errors caused by these systems, such as incorrect diagnoses or treatments, can have severe and even life-threatening consequences for patients. To address this issue, Explainable Artificial Intelligence (XAI) has emerged as a popular area of research, focused on understanding the black-box nature of complex and hard-to-interpret machine learning models. While humans can increase the accuracy of these models through technical expertise, understanding how these models actually function during training can be difficult or even impossible. XAI algorithms such as Local Interpretable Model-Agnostic Explanations (LIME) and SHapley Additive exPlanations (SHAP) can provide explanations for these models, improving trust in their predictions by providing feature importance and increasing confidence in the systems. Many articles have been published that propose solutions to medical problems by using machine learning models alongside XAI algorithms to provide interpretability and explainability. In our study, we identified 454 articles published from 2018–2022 and analyzed 93 of them to explore the use of these techniques in the medical domain.}, journal={Computers in Biology and Medicine}, author={Ali, Subhan and Akhlaq, Filza and Imran, Ali Shariq and Kastrati, Zenun and Daudpota, Sher Muhammad and Moosa, Muhammad}, year={2023}, month=nov, pages={107555} }

@ARTICLE{PubMed_Smak_Gregoor,
  title    = "An artificial intelligence based app for skin cancer detection
              evaluated in a population based setting",
  author   = "Smak Gregoor, Anna M and Sangers, Tobias E and Bakker, Lytske J
              and Hollestein, Loes and Uyl -- de Groot, Carin A and Nijsten,
              Tamar and Wakkee, Marlies",
  journal  = "npj Digital Medicine",
  volume   =  6,
  number   =  1,
  pages    = "90",
  month    =  may,
  year     =  2023
}

@ARTICLE{PubMed_Soliman,
  title    = "Interdisciplinary {Human-Centered} {AI} for Hospital Readmission
              Prediction of Heart Failure Patients",
  author   = "Soliman, Amira and Nair, Monika and Petersson, Marcus and
              Lundgren, Lina and Dryselius, Petra and Fogelberg, Ebba and
              Hamed, Omar and Etminani, Kobra and Nygren, Jens",
  journal  = "Stud Health Technol Inform",
  volume   =  302,
  pages    = "556--560",
  month    =  may,
  year     =  2023,
  address  = "Netherlands",
  language = "en"
}

@book{RAND_Heins,
author="Heins, Sara E. and Denis Agniel and Jacob Mann and Melony E. Sorbero",
title="Reviewing, Refining, and Validating Claims-Based Algorithms of Frailty and Functional Impairment: Final Report",
address="Santa Monica, CA",
year="2023",
doi="10.7249/RRA1871-1",
publisher="RAND Corporation"
}

@article{RAND_Kapinos,
    author = {Kapinos, Kandice A. and Peters, Richard M., Jr and Murphy, Robert E. and Hohmann, Samuel F. and Podichetty, Ankita and Greenberg, Raymond S.},
    title = "{Inpatient Costs of Treating Patients With COVID-19}",
    journal = {JAMA Network Open},
    volume = {7},
    number = {1},
    pages = {e2350145-e2350145},
    year = {2024},
    month = {01},
    issn = {2574-3805},
    doi = {10.1001/jamanetworkopen.2023.50145},
    url = {https://doi.org/10.1001/jamanetworkopen.2023.50145},
}

@ARTICLE{Springer_Kamari,
  title    = "Clinical performance of {AI-integrated} risk assessment pooling
              reveals cost savings even at high prevalence of {COVID-19}",
  author   = "Kamari, Farzin and Eller, Esben and B{\o}gebjerg, Mathias Emil
              and Capella, Ignacio Mart{\'\i}nez and Galende, Borja Arroyo and
              Korim, Tomas and {\O}land, Pernille and Borup, Martin Lysbjerg
              and Frederiksen, Anja R{\aa}dberg and Ranjouriheravi, Amir and
              Al-Jwadi, Ahmed Faris and Mansour, Mostafa and Hansen, Sara and
              Diethelm, Isabella and Burek, Marta and Alvarez, Federico and
              Buch, Anders Glent and Mojtahedi, Nima and R{\"o}ttger, Richard
              and Segtnan, Eivind Antonsen",
  journal  = "Scientific Reports",
  volume   =  14,
  number   =  1,
  pages    = "8853",
  month    =  apr,
  year     =  2024
}

@ARTICLE{Elsevier_Davoudi,
  title    = "Fairness gaps in Machine learning models for hospitalization and
              emergency department visit risk prediction in home healthcare
              patients with heart failure",
  author   = "Davoudi, Anahita and Chae, Sena and Evans, Lauren and Sridharan,
              Sridevi and Song, Jiyoun and Bowles, Kathryn H and McDonald,
              Margaret V and Topaz, Maxim",
  journal  = "International Journal of Medical Informatics",
  pages    = "105534",
  month    =  jun,
  year     =  2024
}

@ARTICLE{Elsevier_Gentili,
  title    = "Machine learning from real data: A mental health registry case
              study",
  author   = "Gentili, Elisabetta and Franchini, Giorgia and Zese, Riccardo and
              Alberti, Marco and Ferrara, Maria and Domenicano, Ilaria and
              Grassi, Luigi",
  journal  = "Computer Methods and Programs in Biomedicine Update",
  volume   =  5,
  pages    = "100132",
  month    =  jan,
  year     =  2024
}

@ARTICLE{Elsevier_Haeberle_Koch,
  title    = "Artificial {Intelligence-Based} Preeclampsia Prediction and Cost
              Benefits for the Health Care System in the United States",
  author   = "Haeberle, Marvin and Koch, Gilbert",
  journal  = "Journal of Obstetric, Gynecologic \& Neonatal Nursing",
  volume   =  53,
  number   = "4, Supplement",
  pages    = "S84",
  month    =  may,
  year     =  2024
}

@ARTICLE{Elsevier_Ismukhamedova,
  title    = "Integrating machine learning in electronic health passport based
              on {WHO} study and healthcare resources",
  author   = "Ismukhamedova, Aigerim and Uvaliyeva, Indira and Belginova, Saule",
  journal  = "Informatics in Medicine Unlocked",
  volume   =  44,
  pages    = "101428",
  month    =  jan,
  year     =  2024
}

@ARTICLE{Elsevier_Kanwal,
  title    = "Diagnosis of {Community-Acquired} pneumonia in children using
              photoplethysmography and Machine learning-based classifier",
  author   = "Kanwal, Kehkashan and Khalid, Syed Ghufran and Asif, Muhammad and
              Zafar, Farhana and Qurashi, Aisha Ghazal",
  journal  = "Biomedical Signal Processing and Control",
  volume   =  87,
  pages    = "105367",
  month    =  jan,
  year     =  2024
}

@ARTICLE{Elsevier_Li,
  title    = "Automating and improving cardiovascular disease prediction using
              Machine learning and {EMR} data features from a regional
              healthcare system",
  author   = "Li, Qi and Campan, Alina and Ren, Ai and Eid, Wael E",
  journal  = "International Journal of Medical Informatics",
  volume   =  163,
  pages    = "104786",
  month    =  jul,
  year     =  2022
}

@ARTICLE{Elsevier_Lu,
  title    = "Identifying modifiable and nonmodifiable cost drivers of
              ambulatory rotator cuff repair: a machine learning analysis",
  author   = "Lu, Yining and Labott, Joshua R and Salmons, IV, Harold I and
              Gross, Benjamin D and Barlow, Jonathan D and Sanchez-Sotelo,
              Joaquin and Camp, Christopher L",
  journal  = "Journal of Shoulder and Elbow Surgery",
  volume   =  31,
  number   =  11,
  pages    = "2262--2273",
  month    =  nov,
  year     =  2022
}

@ARTICLE{Elsevier_Meng,
  title    = "Health care utilization and potentially preventable adverse
              outcomes of high-need, high-cost middle-aged and older adults:
              Needs for integrated care models with life-course approach",
  author   = "Meng, Lin-Chieh and Huang, Shih-Tsung and Chen, Ho-Min and
              Hashmi, Ardeshir Z and Hsiao, Fei-Yuan and Chen, Liang-Kung",
  journal  = "Archives of Gerontology and Geriatrics",
  volume   =  109,
  pages    = "104956",
  month    =  jun,
  year     =  2023
}

@ARTICLE{Elsevier_Mohanty,
  title    = "Machine learning for predicting readmission risk among the frail:
              Explainable {AI} for healthcare",
  author   = "Mohanty, Somya D and Lekan, Deborah and McCoy, Thomas P and
              Jenkins, Marjorie and Manda, Prashanti",
  journal  = "Patterns",
  volume   =  3,
  number   =  1,
  pages    = "100395",
  month    =  jan,
  year     =  2022
}

@ARTICLE{Elsevier_Orji_Ukwandu,
  title    = "Machine learning for an explainable cost prediction of medical
              insurance",
  author   = "Orji, Ugochukwu and Ukwandu, Elochukwu",
  journal  = "Machine Learning with Applications",
  volume   =  15,
  pages    = "100516",
  month    =  mar,
  year     =  2024
}

@ARTICLE{Elsevier_Salmons_1,
  title    = "Identifying Modifiable Cost Drivers of Outpatient
              Unicompartmental Knee Arthroplasty With Machine Learning",
  author   = "Salmons, Harold I and Lu, Yining and Labott, Joshua R and Wyles,
              Cody C and Camp, Christopher L and Taunton, Michael J",
  journal  = "The Journal of Arthroplasty",
  volume   =  38,
  number   =  10,
  pages    = "2051--2059.e2",
  month    =  oct,
  year     =  2023
}

@ARTICLE{Elsevier_Salmons_2,
  title    = "Implementation of Machine Learning to Predict Cost of Care
              Associated with Ambulatory {Single-Level} Lumbar Decompression",
  author   = "Salmons, Harold I and Lu, Yining and Reed, Ryder R and Forsythe,
              Brian and Sebastian, Arjun S",
  journal  = "World Neurosurgery",
  volume   =  167,
  pages    = "e1072--e1079",
  month    =  nov,
  year     =  2022
}

@ARTICLE{Elsevier_Sanderson,
  title    = "Predicting death by suicide following an emergency department
              visit for parasuicide with administrative health care system data
              and machine learning",
  author   = "Sanderson, Michael and Bulloch, Andrew G M and Wang, Jianli and
              Williams, Kimberly G and Williamson, Tyler and Patten, Scott B",
  journal  = "EClinicalMedicine",
  volume   =  20,
  pages    = "100281",
  month    =  mar,
  year     =  2020
}

@ARTICLE{Elsevier_Wei,
  title    = "Machine learning to understand risks for severe {COVID-19}
              outcomes: a retrospective cohort study of immune-mediated
              inflammatory diseases, immunomodulatory medications, and
              comorbidities in a large {US} health-care system",
  author   = "Wei, Qi and Mease, Philip J and Chiorean, Michael and Iles-Shih,
              Lulu and Matos, Wanessa F and Baumgartner, Andrew and Molani,
              Sevda and Hwang, Yeon Mi and Belhu, Basazin and Ralevski,
              Alexandra and Hadlock, Jennifer",
  journal  = "The Lancet Digital Health",
  volume   =  6,
  number   =  5,
  pages    = "e309--e322",
  month    =  may,
  year     =  2024
}

@ARTICLE{Elsevier_Zea-Vera,
  title    = "Machine Learning to Predict Outcomes and Cost by Phase of Care
              After Coronary Artery Bypass Grafting",
  author   = "Zea-Vera, Rodrigo and Ryan, Christopher T and Havelka, Jim and
              Corr, Stuart J and Nguyen, Tom C and Chatterjee, Subhasis and
              Wall, Matthew J and Coselli, Joseph S and Rosengart, Todd K and
              Ghanta, Ravi K",
  journal  = "The Annals of Thoracic Surgery",
  volume   =  114,
  number   =  3,
  pages    = "711--719",
  month    =  sep,
  year     =  2022
}

@inproceedings{home_health,
author = {Wu, Yiying and Lee, Jung-Joo and Pillai, Ajit G. and Cho, Janghee and Ahmadpour, Naseem and Roto, Virpi and Sachathep, Thida and Liu, Jiashuo and Sawan, Mouna and Song, Dongjin and \v{C}ai\'{c}, Martina and Cheng, Lucas and Liu, Renxuan and Kettley, Sarah and Soares, Luis and Grace, Kazjon and Astell-Burt, Thomas},
title = {Collective Imaginaries for the Futures of Care Work},
year = {2024},
isbn = {9798400711145},
publisher = {Association for Computing Machinery},
address = {New York, NY, USA},
url = {https://doi.org/10.1145/3678884.3681838},
doi = {10.1145/3678884.3681838},
booktitle = {Companion Publication of the 2024 Conference on Computer-Supported Cooperative Work and Social Computing},
pages = {732–735},
numpages = {4},
location = {San Jose, Costa Rica},
series = {CSCW Companion '24}
}

@inproceedings{haque_RR, address={New York, NY, USA}, series={CHI ’24}, title={Are We Asking the Right Questions?: Designing for Community Stakeholders’ Interactions with AI in Policing}, ISBN={9798400703300}, url={https://dl.acm.org/doi/10.1145/3613904.3642738}, DOI={10.1145/3613904.3642738}, abstractNote={Research into recidivism risk prediction in the criminal justice system has garnered significant attention from HCI, critical algorithm studies, and the emerging field of human-AI decision-making. This study focuses on algorithmic crime mapping, a prevalent yet underexplored form of algorithmic decision support (ADS) in this context. We conducted experiments and follow-up interviews with 60 participants, including community members, technical experts, and law enforcement agents (LEAs), to explore how lived experiences, technical knowledge, and domain expertise shape interactions with the ADS, impacting human-AI decision-making. Surprisingly, we found that domain experts (LEAs) often exhibited anchoring bias, readily accepting and engaging with the first crime map presented to them. Conversely, community members and technical experts were more inclined to engage with the tool, adjust controls, and generate different maps. Our findings highlight that all three stakeholders were able to provide critical feedback regarding AI design and use - community members questioned the core motivation of the tool, technical experts drew attention to the elastic nature of data science practice, and LEAs suggested redesign pathways such that the tool could complement their domain expertise.}, booktitle={Proceedings of the 2024 CHI Conference on Human Factors in Computing Systems}, publisher={Association for Computing Machinery}, author={Haque, MD Romael and Saxena, Devansh and Weathington, Katy and Chudzik, Joseph and Guha, Shion}, year={2024}, month=may, pages={1–20}, collection={CHI ’24} }

@inproceedings{devansh_group,
author = {Saxena, Devansh},
title = {Designing Human-Centered Algorithms for the Public Sector A Case Study of the U.S. Child-Welfare System},
year = {2023},
isbn = {9781450399456},
publisher = {Association for Computing Machinery},
address = {New York, NY, USA},
url = {https://doi.org/10.1145/3565967.3571759},
doi = {10.1145/3565967.3571759},
booktitle = {Companion Proceedings of the 2023 ACM International Conference on Supporting Group Work},
pages = {66–68},
numpages = {3},
location = {Hilton Head, SC, USA},
series = {GROUP '23}
}

@inproceedings{saxena_risk,
author = {Saxena, Devansh and Moon, Erina Seh-Young and Chaurasia, Aryan and Guan, Yixin and Guha, Shion},
title = {Rethinking "Risk" in Algorithmic Systems Through A Computational Narrative Analysis of Casenotes in Child-Welfare},
year = {2023},
isbn = {9781450394215},
publisher = {Association for Computing Machinery},
address = {New York, NY, USA},
url = {https://doi.org/10.1145/3544548.3581308},
doi = {10.1145/3544548.3581308},
booktitle = {Proceedings of the 2023 CHI Conference on Human Factors in Computing Systems},
articleno = {873},
numpages = {19},
location = {Hamburg, Germany},
series = {CHI '23}
}

@inproceedings{Saxena2022,
author = {Saxena, Devansh and Moon, Seh Young and Shehata, Dahlia and Guha, Shion},
title = {Unpacking Invisible Work Practices, Constraints, and Latent Power Relationships in Child Welfare through Casenote Analysis},
year = {2022},
isbn = {9781450391573},
publisher = {Association for Computing Machinery},
address = {New York, NY, USA},
url = {https://doi.org/10.1145/3491102.3517742},
doi = {10.1145/3491102.3517742},
booktitle = {Proceedings of the 2022 CHI Conference on Human Factors in Computing Systems},
articleno = {120},
numpages = {22},
location = {New Orleans, LA, USA},
series = {CHI '22}
}

@inproceedings{Moon2024,
author = {Moon, Erina Seh-Young and Guha, Shion},
title = {A Human-Centered Review of Algorithms in Homelessness Research},
year = {2024},
isbn = {9798400703300},
publisher = {Association for Computing Machinery},
address = {New York, NY, USA},
url = {https://doi.org/10.1145/3613904.3642392},
doi = {10.1145/3613904.3642392},
booktitle = {Proceedings of the CHI Conference on Human Factors in Computing Systems},
articleno = {870},
numpages = {15},
location = {Honolulu, HI, USA},
series = {CHI '24}
}

@article{pam_review_2,
author = {Razi, Afsaneh and Kim, Seunghyun and Alsoubai, Ashwaq and Stringhini, Gianluca and Solorio, Thamar and De Choudhury, Munmun and Wisniewski, Pamela J.},
title = {A Human-Centered Systematic Literature Review of the Computational Approaches for Online Sexual Risk Detection},
year = {2021},
issue_date = {October 2021},
publisher = {Association for Computing Machinery},
address = {New York, NY, USA},
volume = {5},
number = {CSCW2},
url = {https://doi.org/10.1145/3479609},
doi = {10.1145/3479609},
journal = {Proc. ACM Hum.-Comput. Interact.},
month = oct,
articleno = {465},
numpages = {38}
}

@article{pam_review_1,
author = {Kim, Seunghyun and Razi, Afsaneh and Stringhini, Gianluca and Wisniewski, Pamela J. and De Choudhury, Munmun},
title = {A Human-Centered Systematic Literature Review of Cyberbullying Detection Algorithms},
year = {2021},
issue_date = {October 2021},
publisher = {Association for Computing Machinery},
address = {New York, NY, USA},
volume = {5},
number = {CSCW2},
url = {https://doi.org/10.1145/3476066},
doi = {10.1145/3476066},
journal = {Proc. ACM Hum.-Comput. Interact.},
month = oct,
articleno = {325},
numpages = {34}
}

@inproceedings{saxena_lit,
author = {Saxena, Devansh and Badillo-Urquiola, Karla and Wisniewski, Pamela J. and Guha, Shion},
title = {A Human-Centered Review of Algorithms used within the U.S. Child Welfare System},
year = {2020},
isbn = {9781450367080},
publisher = {Association for Computing Machinery},
address = {New York, NY, USA},
url = {https://doi.org/10.1145/3313831.3376229},
doi = {10.1145/3313831.3376229},
booktitle = {Proceedings of the 2020 CHI Conference on Human Factors in Computing Systems},
pages = {1–15},
numpages = {15},
location = {Honolulu, HI, USA},
series = {CHI '20}
}

@inproceedings{mcconvey2023,
author = {McConvey, Kelly and Guha, Shion and Kuzminykh, Anastasia},
title = {A Human-Centered Review of Algorithms in Decision-Making in Higher Education},
year = {2023},
isbn = {9781450394215},
publisher = {Association for Computing Machinery},
address = {New York, NY, USA},
url = {https://doi.org/10.1145/3544548.3580658},
doi = {10.1145/3544548.3580658},
booktitle = {Proceedings of the 2023 CHI Conference on Human Factors in Computing Systems},
articleno = {223},
numpages = {15},
location = {Hamburg, Germany},
series = {CHI '23}
}

@inproceedings{chadli,
author = {Chadli, Kouider and Botterweck, Goetz and Saber, Takfarinas},
title = {The Environmental Cost of Engineering Machine Learning-Enabled Systems: A Mapping Study},
year = {2024},
isbn = {9798400705410},
publisher = {Association for Computing Machinery},
address = {New York, NY, USA},
url = {https://doi.org/10.1145/3642970.3655828},
doi = {10.1145/3642970.3655828},
booktitle = {Proceedings of the 4th Workshop on Machine Learning and Systems},
pages = {200–207},
numpages = {8},
location = {Athens, Greece},
series = {EuroMLSys '24}
}

@inproceedings{McConvey2024,
author = {McConvey, Kelly and Guha, Shion},
title = {"This is not a data problem": Algorithms and Power in Public Higher Education in Canada},
year = {2024},
isbn = {9798400703300},
publisher = {Association for Computing Machinery},
address = {New York, NY, USA},
url = {https://doi.org/10.1145/3613904.3642451},
doi = {10.1145/3613904.3642451},
booktitle = {Proceedings of the CHI Conference on Human Factors in Computing Systems},
articleno = {16},
numpages = {14},
location = {Honolulu, HI, USA},
series = {CHI '24}
}

@inproceedings{Chui2023,
author = {Chui, Victoria and Pater, Jessica and Toscos, Tammy and Guha, Shion},
title = {Applying Human-Centered Data Science to Healthcare: Hyperlocal Modeling of COVID-19 Hospitalizations},
year = {2023},
isbn = {9781450399456},
publisher = {Association for Computing Machinery},
address = {New York, NY, USA},
url = {https://doi.org/10.1145/3565967.3570979},
doi = {10.1145/3565967.3570979},
booktitle = {Companion Proceedings of the 2023 ACM International Conference on Supporting Group Work},
pages = {24–26},
numpages = {3},
location = {Hilton Head, SC, USA},
series = {GROUP '23}
}

@ARTICLE{Erion2022,
  title    = "A cost-aware framework for the development of {AI} models for
              healthcare applications",
  author   = "Erion, Gabriel and Janizek, Jose and Hudelson, Carly and
              Utarnachitt, Richard B and McCoy, Andrew M and Sayre, Michael R
              and White, Nathan J and Lee, Su-In",
  journal  = "Nature Biomedical Engineering",
  volume   =  6,
  number   =  12,
  pages    = "1384--1398",
  month    =  dec,
  year     =  2022
}

@article{AI_Personalized_Gifari,
	title = {Artificial {Intelligence} toward {Personalized} {Medicine}},
	volume = {8},
	issn = {2477-0612},
	url = {https://scholarhub.ui.ac.id/psr/vol8/iss2/1},
	doi = {10.7454/psr.v8i2.1199},
	number = {2},
	journal = {Pharmaceutical Sciences and Research},
	author = {Gifari, Muhammad and Samodro, Pugud and Kurniawan, Dhadhang},
	month = aug,
	year = {2021},
}

@article{Williams_NLP_unstructured, title={Natural Language Processing for Unlocking Insights from Unstructured Big Data in The Healthcare Industry}, volume={15}, url={https://orientreview.com/index.php/etmibd-journal/article/view/27}, abstractNote={&amp;lt;p&amp;gt;Healthcare&amp;#039;s vast data volume is rapidly growing, of which over 80% is unstructured clinical notes, medical images, literature publications and social conversations. This big text data hides invaluable insights to enhance decisions, outcomes and discoveries. Natural language processing (NLP) enables extracting value from narratives using linguistics understanding to automatically convert free-text to structured data. This paper discusses NLP techniques applied in healthcare and practical benefits achieved. For electronic health records, dictionary and machine learning entity extraction accurately identifies clinical concepts like symptoms and treatments in notes for decision support, while relation extraction reveals links between medical problems and medications improving clinical modeling. Summarization of lengthy records also assists physicians. On social platforms, NLP helps accelerating discoveries by uncovering public health insights around outbreak forecasting, adverse drug events monitoring, and mental health conditions absent in curated medical datasets. For biomedicine&amp;#039;s unstructured knowledge in publications, machine reading comprehension enables hypothesis generation and validation by answering complex questions with over 90% accuracy. While accuracy, security and interoperability challenges persist, innovations in transfer learning from language models like BERT, graph-based contextual representation, user-centered design, and federated learning are overcoming adoption barriers. As ethical implications are addressed responsibly, NLP adoption is expected to rise steeply. Overall, NLP unlocks healthcare&amp;#039;s big unstructured data, unlocking evidence and insights supporting improved clinical and operational outcomes, patient-centric care and transformative medical discoveries using AI techniques that perceive both the content and contexts encoded in natural language.&amp;lt;/p&amp;gt;}, number={10}, journal={Emerging Trends in Machine Intelligence and Big Data}, author={Williams, Samantha and Petrovich, Elena}, year={2023}, month={Oct.}, pages={30–39} }

@ARTICLE{Sedlakova_unstructured_narratives,
  title    = "Challenges and best practices for digital unstructured data
              enrichment in health research: A systematic narrative review",
  author   = "Sedlakova, Jana and Daniore, Paola and Horn Wintsch, Andrea and
              Wolf, Markus and Stanikic, Mina and Haag, Christina and Sieber,
              Chlo{\'e} and Schneider, Gerold and Staub, Kaspar and Alois
              Ettlin, Dominik and Gr{\"u}bner, Oliver and Rinaldi, Fabio and
              von Wyl, Viktor and {University of Zurich Digital Society
              Initiative (UZH-DSI) Health Community}",
  journal  = "PLOS Digit Health",
  volume   =  2,
  number   =  10,
  pages    = "e0000347",
  month    =  oct,
  year     =  2023,
  address  = "United States",
  language = "en"
}

@article{personalized_AI_Johnson,
author = {Johnson, Kevin B. and Wei, Wei-Qi and Weeraratne, Dilhan and Frisse, Mark E. and Misulis, Karl and Rhee, Kyu and Zhao, Juan and Snowdon, Jane L.},
title = {Precision Medicine, AI, and the Future of Personalized Health Care},
journal = {Clinical and Translational Science},
volume = {14},
number = {1},
pages = {86-93},
doi = {https://doi.org/10.1111/cts.12884},
url = {https://ascpt.onlinelibrary.wiley.com/doi/abs/10.1111/cts.12884},
year = {2021}
}

@article{personalized_AI_costs_Ahmed,
    author = {Ahmed, Zeeshan and Mohamed, Khalid and Zeeshan, Saman and Dong, XinQi},
    title = "{Artificial intelligence with multi-functional machine learning platform development for better healthcare and precision medicine}",
    journal = {Database},
    volume = {2020},
    pages = {baaa010},
    year = {2020},
    month = {03},
    issn = {1758-0463},
    doi = {10.1093/database/baaa010},
    url = {https://doi.org/10.1093/database/baaa010},
    eprint = {https://academic.oup.com/database/article-pdf/doi/10.1093/database/baaa010/32923892/baaa010.pdf},
}

@ARTICLE{Xiao_automate,
  title    = "Health care cost and benefits of artificial intelligence-assisted
              population-based glaucoma screening for the elderly in remote
              areas of China: a cost-offset analysis",
  author   = "Xiao, Xuan and Xue, Long and Ye, Lin and Li, Hongzheng and He,
              Yunzhen",
  journal  = "BMC Public Health",
  volume   =  21,
  number   =  1,
  pages    = "1065",
  month    =  jun,
  year     =  2021
}

@article{wojcik_discrimination,
	title = {Algorithmic {Discrimination} in {Health} {Care}},
	volume = {24},
	issn = {1079-0969},
	url = {https://www.ncbi.nlm.nih.gov/pmc/articles/PMC9212826/},
	number = {1},
	urldate = {2024-07-29},
	journal = {Health and Human Rights},
	author = {Wójcik, Malwina Anna},
	month = jun,
	year = {2022},
	pmid = {35747275},
	pmcid = {PMC9212826},
	pages = {93--103},
}

@ARTICLE{digital_medicine,
  title    = "What is digital medicine?",
  author   = "Shaffer, David Williamson and Kigin, Colleen M and Kaput, James J
              and Gazelle, G Scott",
  journal  = "Stud Health Technol Inform",
  volume   =  80,
  pages    = "195--204",
  year     =  2002,
  address  = "Netherlands",
  language = "en"
}

@article{giovanola_bias,
	title = {Beyond bias and discrimination: redefining the {AI} ethics principle of fairness in healthcare machine-learning algorithms},
	volume = {38},
	issn = {1435-5655},
	shorttitle = {Beyond bias and discrimination},
	url = {https://doi.org/10.1007/s00146-022-01455-6},
	doi = {10.1007/s00146-022-01455-6},
	language = {en},
	number = {2},
	urldate = {2024-07-29},
	journal = {AI \& SOCIETY},
	author = {Giovanola, Benedetta and Tiribelli, Simona},
	month = apr,
	year = {2023},
	pages = {549--563},
}

@ARTICLE{e-health,
  title    = "The Limitations of User-and {Human-Centered} Design in an eHealth
              Context and How to Move Beyond Them",
  author   = "van Velsen, Lex and Ludden, Geke and Gr{\"u}nloh, Christiane",
  journal  = "J Med Internet Res",
  volume   =  24,
  number   =  10,
  pages    = "e37341",
  month    =  oct,
  year     =  2022
}

@article{chen_security,
	title = {Generative {AI} in {Medical} {Practice}: {In}-{Depth} {Exploration} of {Privacy} and {Security} {Challenges}},
	volume = {26},
	shorttitle = {Generative {AI} in {Medical} {Practice}},
	url = {https://www.jmir.org/2024/1/e53008},
	doi = {10.2196/53008},
	language = {EN},
	number = {1},
	urldate = {2024-07-29},
	journal = {Journal of Medical Internet Research},
	author = {Chen, Yan and Esmaeilzadeh, Pouyan},
	month = mar,
	year = {2024},
	note = {Company: Journal of Medical Internet Research
Distributor: Journal of Medical Internet Research
Institution: Journal of Medical Internet Research
Label: Journal of Medical Internet Research
Publisher: JMIR Publications Inc., Toronto, Canada},
	pages = {e53008},
}

@ARTICLE{ting,
  title    = "Gaps and future of human-centered artificial intelligence in
              ophthalmology: Future Vision Forum consensus statement",
  author   = "Ting, Daniel Shu Wei and Humayun, Mark S and Huang, Suber S and
              {Future Vision Forum Faculty}",
  journal  = "Current Opinion in Ophthalmology",
  volume   =  34,
  number   =  5,
  pages    = "431--436",
  month    =  sep,
  year     =  2023
}

@ARTICLE{thieme,
  title    = "Machine Learning in Mental Health: A Systematic Review of the
              {HCI} Literature to Support the Development of Effective and
              Implementable {ML} Systems",
  author   = "Thieme, Anja and Belgrave, Danielle and Doherty, Gavin",
  journal  = "ACM Transactions on Computer-Human Interaction",
  volume   =  27,
  number   =  5,
  pages    = "34:1--34:53",
  month    =  aug,
  year     =  2020
}

@ARTICLE{rwanda,
  title    = "Using {Human-Centered} Design to Develop, Launch, and Evaluate a
              National Digital Health Platform to Improve Reproductive Health
              for Rwandan Youth",
  author   = "Ippoliti, Nicole and Sekamana, Mireille and Baringer, Laura and
              Hope, Rebecca",
  journal  = "GLOB HEALTH SCI PRACT",
  volume   =  9,
  number   = "Supplement 2",
  pages    = "S244",
  month    =  nov,
  year     =  2021
}

@article{smart_health,
author = {Neumann, Sara and Bleja, Jelena and Kr\"{u}ger, Tim and Grossmann, Uwe},
title = {Participating Citizens = Smart Citizens? Applying the Human-centered Design Approach on a Digital Care Platform},
year = {2023},
issue_date = {September 2023},
publisher = {Association for Computing Machinery},
address = {New York, NY, USA},
volume = {4},
number = {3},
url = {https://doi.org/10.1145/3604618},
doi = {10.1145/3604618},
journal = {Digit. Gov.: Res. Pract.},
month = {sep},
articleno = {14},
numpages = {13}
}

@article{obermeyer_bias,
	title = {Dissecting racial bias in an algorithm used to manage the health of populations},
	volume = {366},
	issn = {1095-9203},
	doi = {10.1126/science.aax2342},
	language = {eng},
	number = {6464},
	journal = {Science (New York, N.Y.)},
	author = {Obermeyer, Ziad and Powers, Brian and Vogeli, Christine and Mullainathan, Sendhil},
	month = oct,
	year = {2019},
	pmid = {31649194},
	pages = {447--453},
}

@article{sangers_bias,
	title = {Towards successful implementation of artificial intelligence in skin cancer care: a qualitative study exploring the views of dermatologists and general practitioners},
	volume = {315},
	issn = {1432-069X},
	shorttitle = {Towards successful implementation of artificial intelligence in skin cancer care},
	url = {https://doi.org/10.1007/s00403-022-02492-3},
	doi = {10.1007/s00403-022-02492-3},
	language = {en},
	number = {5},
	urldate = {2024-07-29},
	journal = {Archives of Dermatological Research},
	author = {Sangers, Tobias E. and Wakkee, Marlies and Moolenburgh, Folkert J. and Nijsten, Tamar and Lugtenberg, Marjolein},
	month = jul,
	year = {2023},
	pages = {1187--1195},
}

@article{ledford_bias,
	title = {Millions of black people affected by racial bias in health-care algorithms},
	volume = {574},
	copyright = {2021 Nature},
	url = {https://www-nature-com.myaccess.library.utoronto.ca/articles/d41586-019-03228-6},
	doi = {10.1038/d41586-019-03228-6},
	language = {en},
	number = {7780},
	urldate = {2024-07-29},
	journal = {Nature},
	author = {Ledford, Heidi},
	month = oct,
	year = {2019},
	note = {Bandiera\_abtest: a
Cg\_type: News
Publisher: Nature Publishing Group
Subject\_term: Computer science, Health care, Policy, Society},
	pages = {608--609},
}

@inproceedings{barnard_fairness,
	address = {New York, NY, USA},
	series = {{TAS} '23},
	title = {{MACAIF}: {Machine} {Learning} {Auditing} for {Clinical} {AI} {Fairness}},
	isbn = {9798400707346},
	shorttitle = {{MACAIF}},
	url = {https://dl.acm.org/doi/10.1145/3597512.3597522},
	doi = {10.1145/3597512.3597522},
	urldate = {2024-05-30},
	booktitle = {Proceedings of the {First} {International} {Symposium} on {Trustworthy} {Autonomous} {Systems}},
	publisher = {Association for Computing Machinery},
	author = {Barnard, Pepita and Bautista, John Robert and Krook, Joshua and Liu, Anqi and Menéndez, Héctor and Schmidt, Aurora and Sookoor, Tamim},
	month = jul,
	year = {2023},
	pages = {1--4},
}

@article{hort_bias,
author = {Hort, Max and Chen, Zhenpeng and Zhang, Jie M. and Harman, Mark and Sarro, Federica},
title = {Bias Mitigation for Machine Learning Classifiers: A Comprehensive Survey},
year = {2024},
issue_date = {June 2024},
publisher = {Association for Computing Machinery},
address = {New York, NY, USA},
volume = {1},
number = {2},
url = {https://doi.org/10.1145/3631326},
doi = {10.1145/3631326},
journal = {ACM J. Responsib. Comput.},
month = {Jun},
articleno = {11},
numpages = {52}
}

@ARTICLE{nursing,
  title   = "How artificial intelligence is changing nursing",
  author  = "Robert, Nancy",
  journal = "Nursing Management",
  volume  =  50,
  number  =  9,
  year    =  2019
}
\nocite{*}


\newpage

\begin{appendix}

\section{Search Queries} \label{appendix:terms}

\begin{table}[h]
\centering
\renewcommand{\arraystretch}{1.2}
\resizebox{\textwidth}{!}{
\begin{tabular}{lll}
\textbf{Database}         & \textbf{Search Queries}                                                                                            & \textbf{\# of Papers} \\ \hline
\multirow{5}{*}{ACM}      & "artificial intelligence" AND ("healthcare" OR "health care") AND "cost" AND NOT title:"review" AND NOT title:"literature" & \multirow{5}{*}{30}   \\
                          & "AI" AND ("healthcare" OR "health care") AND "cost"                                                                &                       \\
                          & "machine learning" AND ("healthcare" OR "health care") AND "cost" AND NOT title:"review" AND NOT title:"literature"        &                       \\
                          & "machine learning" AND ("healthcare" OR "health care") AND "cost"                                                  &                       \\
                          & Through paper reference lists                                                                                      &                       \\ \hline
\multirow{3}{*}{IEEE}     & "artificial intelligence" AND ("healthcare" OR "health care") AND "cost" AND NOT title:"review" AND NOT title:"literature" & \multirow{3}{*}{16}   \\
                          & "AI" AND ("healthcare" OR "health care") AND "cost"                                                                &                       \\
                          & "machine learning" AND ("healthcare" OR "health care") AND "cost"                                                  &                       \\ \hline
\multirow{2}{*}{Springer} & "AI" AND ("healthcare" OR "health care") AND "cost"                                                                & \multirow{2}{*}{11}   \\
                          & "machine learning" AND ("healthcare" OR "health care") AND "cost"                                                  &                       \\ \hline
\multirow{3}{*}{PubMed}   & "AI" AND ("healthcare" OR "health care") AND "cost" AND NOT "review"                                                   & \multirow{3}{*}{12}   \\
                          & "machine learning" AND ("healthcare" OR "health care") AND "cost"                                                  &                       \\
                          & Through paper reference lists                                                                                      &                       \\ \hline
\multirow{3}{*}{JMIR}     & "artificial intelligence" AND ("healthcare" OR "health care") AND "cost" AND NOT title:"review" AND NOT title:"literature" & \multirow{3}{*}{10}   \\
                          & "AI" AND ("healthcare" OR "health care") AND "cost"                                                                &                       \\
                          & "machine learning" AND ("healthcare" OR "health care") AND "cost"                                                  &                       \\ \hline
Elsevier                  & "machine learning" AND ("healthcare" OR "health care") AND "cost" AND NOT title: "review" AND NOT title:"literature"   & 27                    \\ \hline
Other                     & Through paper reference lists                                                                                      & 8                    
\end{tabular}}
\caption{Corpus Breakdown by Database and Search Queries}
\label{table:queries}
\end{table}


\section{PRISMA Diagram} \label{sec:appendix_graph}
\begin{figure}[h]
    \centering
    \includegraphics[scale=0.5]{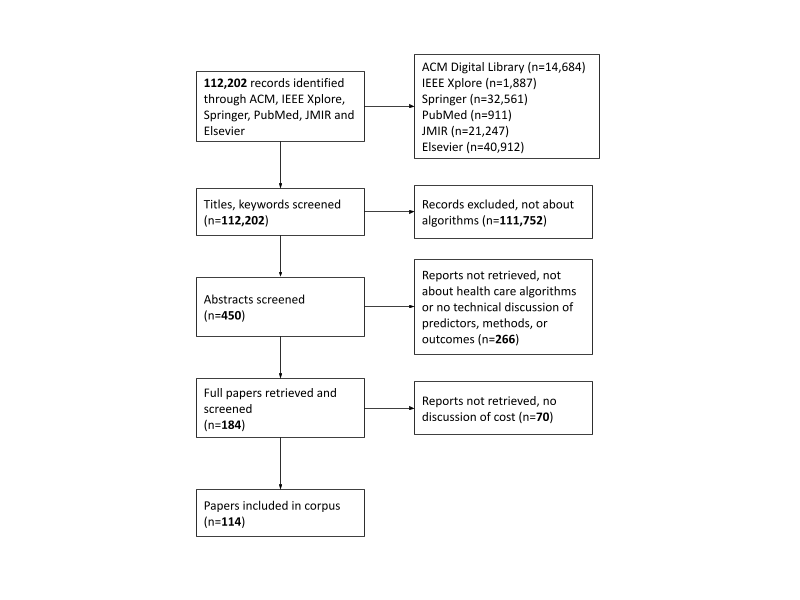}
    \caption{Flow Diagram of Literature Review Process}
    \label{appendix:graph}
\end{figure}

\newpage

\section{Cross-Tabulation Tables} \label{sec:tables_appendix}
\begin{table}[h]
\centering
\resizebox{\textwidth}{!}{\begin{tabular}{l|llll|llll} 
 & \multicolumn{4}{l|}{\textbf{ Method}}                                                                                              & \textbf{Outcome}                                                                         &                                                                                          &                                                                                              &                                                                     \\ \cline{2-9} 
\textbf{Predictor}           & \multicolumn{1}{l|}{\textbf{IS }}
& \multicolumn{1}{l|}{\textbf{IP}} & 
\multicolumn{1}{l|}{\textbf{ML}}
& \textbf{DL} &
\multicolumn{1}{l|}{\textbf{\begin{tabular}[c]{@{}l@{}}Condition \\ Diagnosis \end{tabular}}} & \multicolumn{1}{l|}{\textbf{\begin{tabular}[c]{@{}l@{}}Patient \\ Feature \end{tabular}}} & \multicolumn{1}{l|}{\textbf{\begin{tabular}[c]{@{}l@{}}Risk \\ Prediction\end{tabular}}} & \textbf{\begin{tabular}[c]{@{}l@{}}Cost \\ Prediction \end{tabular}} \\ \hline
\textbf{Demographics}        & 38                                      & 2                                       & 69                                       & 32                 & 23                                                                                           & 19                                                                                       & 11                                                                                       & 11                                                                                       \\
\textbf{Hospital Admissions} & 22                                      & 1                                       & 38                                       & 17                 & 10                                                                                           & 10                                                                                       & 7                                                                                        & 8                                                                                        \\
\textbf{EHR}                 & 19                                      & 2                                       & 37                                       & 14                 & 11                                                                                           & 9                                                                                        & 9                                                                                        & 5                                                                                        \\
\textbf{Laboratory Results}  & 9                                       & 1                                       & 28                                       & 16                 & 16                                                                                           & 5                                                                                        & 3                                                                                        & 0                                                                                        \\
\textbf{Claims Data}         & 12                                      & 0                                       & 20                                       & 10                 & 2                                                                                            & 7                                                                                        & 3                                                                                        & 6                                                                                        \\
\textbf{Physician Notes}     & 3                                       & 1                                       & 16                                       & 11                 & 8                                                                                            & 3                                                                                        & 3                                                                                        & 0                                                                                        \\
\textbf{Prescriptions}       & 10                                      & 0                                       & 15                                       & 8                  & 6                                                                                            & 7                                                                                        & 1                                                                                        & 1                                                                                        \\
\textbf{Images}              & 2                                       & 10                                      & 16                                       & 14                 & 14                                                                                           & 2                                                                                        & 1                                                                                        & 0    
\end{tabular}}
\caption{Cross-tabulation between predictors, outcome variables, and methods. Each paper in the corpus can be multi-counted across predictor, method, and outcome choices.}
\label{table:all_cross_tab}
\end{table}

\begin{table}[h]
\resizebox{\textwidth}{!}{\begin{tabular}{l|lllllll}
                                  & \multicolumn{7}{l}{\textbf{Method}}                                                                                                                                                                                                                                                                                                                                           \\ \cline{2-8} 
                                  & \multicolumn{2}{l|}{\textbf{\begin{tabular}[c]{@{}l@{}}Inferential \\ Statistics\end{tabular}}} & \multicolumn{1}{l|}{\textbf{\begin{tabular}[c]{@{}l@{}}Image \\ Processing\end{tabular}}} & \multicolumn{3}{l|}{\textbf{\begin{tabular}[c]{@{}l@{}}Machine \\ Learning\end{tabular}}}                   & \textbf{\begin{tabular}[c]{@{}l@{}}Deep \\ Learning\end{tabular}} \\ \cline{2-8} 
\textbf{Target Outcome}           & \multicolumn{1}{l|}{\textbf{ST}}               & \multicolumn{1}{l|}{\textbf{GLM}}              & \multicolumn{1}{l|}{\textbf{IP}}                                                          & \multicolumn{1}{l|}{\textbf{SUP}} & \multicolumn{1}{l|}{\textbf{UNSUP}} & \multicolumn{1}{l|}{\textbf{NLP}} & \textbf{NN}                                                       \\ \hline
\textbf{Condition Diagnosis}      & \multicolumn{1}{l|}{5}                         & \multicolumn{1}{l|}{11}                        & \multicolumn{1}{l|}{8}                                                                    & 35                                & 1                                   & \multicolumn{1}{l|}{1}            & 28                                                                \\
\textbf{Patient Feature Analysis} & \multicolumn{1}{l|}{8}                         & \multicolumn{1}{l|}{10}                        & \multicolumn{1}{l|}{1}                                                                    & 24                                & 4                                   & \multicolumn{1}{l|}{3}            & 12                                                                \\
\textbf{Risk Prediction}          & \multicolumn{1}{l|}{4}                         & \multicolumn{1}{l|}{3}                         & \multicolumn{1}{l|}{1}                                                                    & 15                                & 1                                   & \multicolumn{1}{l|}{1}            & 9                                                                 \\
\textbf{Cost Prediction}          & \multicolumn{1}{l|}{3}                         & \multicolumn{1}{l|}{8}                         & \multicolumn{1}{l|}{0}                                                                    & 9                                 & 0                                   & \multicolumn{1}{l|}{0}            & 2                                                                 \\
\textbf{Patient Experience}       & \multicolumn{1}{l|}{2}                         & \multicolumn{1}{l|}{1}                         & \multicolumn{1}{l|}{0}                                                                    & 0                                 & 3                                   & \multicolumn{1}{l|}{1}            & 1                                                                
\end{tabular}}
\caption{Cross-tabulation between outcome variables and methods}
\label{table:cross_tab2}
\end{table}

\begin{table}[t]
\resizebox{\textwidth}{!}{
\begin{tabular}{l|llll}
                        & \multicolumn{1}{l}{\textbf{Condition Diagnosis}} & \multicolumn{1}{l}{\textbf{Patient Characteristic}} & \multicolumn{1}{l}{\textbf{Risk Prediction}} & \multicolumn{1}{l}{\textbf{Cost Prediction}} \\ \hline
\textbf{Computational}      & 11 (29.7)                                               & 4 (14.3)                                                  & 6 (40.0)                                          & 1 (9.0)                                            \\
\textbf{Organizational} & 5 (13.5)                                                & 2 (7.1)                                                   & 0                                            & 0                                            \\
\textbf{Human/Social}   & 1 (2.7)                                               & 0                                                   & 0                                            & 0                                           
\end{tabular}}
\caption{Count (\%) of Model Cost Category by Outcome}
\label{table:cost_breakdown}
\end{table}

\end{appendix}

\end{document}